\documentclass[11pt,a4paper]{article}
\usepackage{amsfonts}
\usepackage{amsmath}
\usepackage{amssymb}
\usepackage{amsthm}
\usepackage{graphicx}
\usepackage{booktabs}
\usepackage{theorem}
\usepackage{array}
\usepackage[numbers,sort]{natbib}
\usepackage[footnotesize]{caption}
\usepackage{colordvi}

\allowdisplaybreaks[1]

\numberwithin{equation}{section}

\newcommand{\be}{\begin{equation}}
\newcommand{\ee}{\end{equation}}
\newcommand{\beq}{\begin{equation}}
\newcommand{\eeq}{\end{equation}}
\newcommand{\bea}{\begin{eqnarray}}
\newcommand{\eea}{\end{eqnarray}}

\newcommand{\ol}{\overline}
\newcommand{\wt}{\widetilde}
\newcommand{\bs}{\boldsymbol}

\makeatletter

\newcommand{\Rmnum}[1]{\expandafter\@slowromancap\romannumeral #1@}
\makeatother

\begin{document}

\begin{titlepage}

\vspace*{-15mm}
\begin{flushright}
MPP-2011-38\\
SHEP-11-04\\
SISSA 15/2011/EP
\end{flushright}
\vspace*{0.7cm}

\begin{center}
{
\bf\LARGE
Right unitarity triangles and tri-bimaximal mixing \\[4mm]
from discrete symmetries and unification}
\\[9mm]
S.~Antusch$^{\star}$
\footnote{E-mail: \texttt{antusch@mppmu.mpg.de}},
Stephen~F.~King$^{\dagger}$
\footnote{E-mail: \texttt{king@soton.ac.uk}},
Christoph~Luhn$^{\dagger}$
\footnote{E-mail: \texttt{christoph.luhn@soton.ac.uk}},
M.~Spinrath$^{\S}$
\footnote{E-mail: \texttt{spinrath@sissa.it}},
\\[-2mm]

\end{center}
\vspace*{0.50cm}
\centerline{$^{\star}$ \it
Max-Planck-Institut f\"ur Physik (Werner-Heisenberg-Institut),}
\centerline{\it
F\"ohringer Ring 6, D-80805 M\"unchen, Germany}
\vspace*{0.2cm}
\centerline{$^{\dagger}$ \it
School of Physics and Astronomy, University of Southampton,}
\centerline{\it
SO17 1BJ Southampton, United Kingdom }
\vspace*{0.2cm}
\centerline{$^{\S}$ \it
SISSA/ISAS and INFN,}
\centerline{\it
Via Bonomea 265, I-34136 Trieste, Italy }
\vspace*{1.20cm}
\begin{abstract}

\noindent
We propose new classes of models which predict both
tri-bimaximal lepton mixing and a right-angled
Cabibbo-Kobayashi-Maskawa (CKM) unitarity triangle, $\alpha \approx 90^\circ$.
The ingredients of the models include a supersymmetric (SUSY) unified gauge group such as $SU(5)$,
a discrete family symmetry such as $A_4$ or $S_4$,
a shaping symmetry including products of $\mathbb{Z}_2$ and $\mathbb{Z}_4$ groups
as well as spontaneous CP violation.
We show how the vacuum alignment in such models allows 
a simple explanation of $\alpha \approx 90^\circ$ by a combination of 
purely real or purely imaginary vacuum expectation values (vevs) of the flavons
responsible for family symmetry breaking. This leads to quark mass matrices
with 1-3 texture zeros that satisfy the ``phase sum rule''
and lepton mass matrices that satisfy the ``lepton mixing sum rule'' together with a new prediction that the 
leptonic CP violating oscillation phase is close to either $0^\circ$, $90^\circ$,
$180^\circ$, or $270^\circ$ depending on the model,
with neutrino masses being purely real (no complex Majorana phases).
This leads to the possibility of having right-angled unitarity triangles in both the quark and lepton sectors.

\end{abstract}

\end{titlepage}

\setcounter{footnote}{0}

\section{Introduction}
The flavour puzzle, i.e.\ the origin of the observed pattern of fermion masses, mixing angles and CP violating phases is one of the most challenging puzzles in particle physics. There are various aspects of the flavour puzzle, such as the hierarchy among the quark masses, the origin of CP violation, and the largeness of the leptonic mixing angles, which have turned out to be close to ``tri-bimaximal'' \cite{HPS}. 
In particular this latter observation has led to increasing interest in
non-Abelian discrete family symmetries for flavour model
building~\cite{Altarelli:2010gt}.

Recently it has become increasingly clear that current data is indeed consistent with the hypothesis of a right-angled CKM unitarity triangle, with the best fits giving $\alpha = \left(89.0^{+4.4}_{-4.2} \right)^\circ$~\cite{PDG}. 
Such a right unitarity triangle was suggested long ago, when the error bar on $\alpha$ was much larger, by Fritzsch and collaborators as a natural consequence
of having quark mass matrices with zeros in the 1-3 element \cite{Fritzsch:1979zq}.
In the light of recent data, this observation has gained increased momentum,
and there have been several papers that attempt to predict $\alpha \approx
90^\circ$ by postulating up-type and down-type quark mass matrices with
the elements $M^{u,d}_{ij}$ being either purely real or purely imaginary, with
texture zeros in the 1-3 elements, $M^{u,d}_{13}=0$
\cite{Masina:2006pe,Xing:2009eg,Antusch:2009hq} (see also \cite{Scott}). 
Under these assumptions, it has been shown that the prediction $\alpha \approx
90^\circ$ can be understood from a simple analytic ``phase sum rule'' relation
\cite{Antusch:2009hq} which relates the angle~$\alpha$ to phases arising from
the quark mass matrices. To be precise, the ``phase sum rule'' can be expressed as $\alpha \approx \delta_{12}^d - \delta_{12}^u$~\cite{Antusch:2009hq}, where the phases $\delta_{12}^d$ and $\delta_{12}^u$ are the arguments of the complex 1-2 rotation angles in the up-type and down-type quark mass matrices, as defined in  \cite{Antusch:2009hq}. To explain $\alpha \approx  90^\circ$ one might therefore simply try to realise $\delta_{12}^d =  90^\circ$, $\delta_{12}^u = 0$ or alternatively $\delta_{12}^d = 0$, $\delta_{12}^u = - 90^\circ$ in a model of flavour. For hierarchical mass matrices, this corresponds to the 1-2 and the 2-2 elements of the mass matrix being either purely real or imaginary. 
Such textures have also been considered previously in  \cite{Masina:2006pe,Xing:2009eg,Antusch:2009hq}, 
and in \cite{Masina:2006pe} a SUSY Grand Unified Theory (GUT) with a continuous family symmetry $SU(3)\times SU(3)$ responsible for such textures has been proposed, however the vacuum alignment responsible for the spontaneous breaking of the family symmetry was not studied.

In this paper we show that large classes of models involving discrete family symmetry and supersymmetric unification
(so called SUSY GUTs of Flavour) that have previously been proposed to account for tri-bimaximal lepton mixing are quite capable of providing an explanation of the right-angled unitarity triangle, subject to a constraint
on the ``shaping symmetry'' that helps to shape the vacuum alignment superpotential. 
Such a ``shaping symmetry'' is always necessary in realistic models, but here we constrain the 
nature of the symmetry to be a discrete symmetry of a particular type. 
The main technical accomplishment in this paper is to propose a mechanism for vacuum alignment based on discrete symmetries, which can give rise to 
purely real or purely imaginary vacuum alignments for the flavon fields
responsible for spontaneously breaking the discrete family
symmetry.\footnote{In the context of continuous family symmetries, flavon
  alignments with real or imaginary values in the multiplet components have
  been discussed in~\cite{Barbieri:1999km}.} There are four different aspects to these models which are important for our approach, as follows.

(i) {\bf Supersymmetric Unification}: We impose gauge coupling unification, which severely restricts the 
choice of available models in the literature, since many of the existing models are not unified.
The role of unification is to relate the lepton sector to the quark sector, since we want to make the
connection between tri-bimaximal mixing and quark mixing and CP violation. Here we shall consider 
the minimal SUSY $SU(5)$ gauge group. It is, however, worth emphasising that
our method of obtaining a right-angled CKM unitarity triangle can also be
applied to models without grand unification.

(ii) {\bf Discrete Family Symmetry}:
As already stated we are also concerned with discrete family symmetries that
have been proposed to account for tri-bimaximal lepton mixing. The approach is applicable for all types of discrete family symmetries and 
does not depend on whether the neutrino flavour (Klein) symmetry associated with tri-bimaximal mixing is identified as a subgroup of the family symmetry (as in the so called direct models) or as an accidental symmetry which results from having flavons aligned along the columns of the tri-bimaximal mixing matrix (as in the so called indirect models). Recall that in the latter case, the flavons break the Klein symmetry only due to an overall minus sign, and bilinears of flavons appearing in the neutrino sector respect the Klein symmetry (for a full discussion of this see \cite{King:2009ap}). We shall consider an example model of both the direct and the indirect kind.

(iii) {\bf Discrete Shaping Symmetry}: 
We assume an extra shaping symmetry based on products of $\mathbb{Z}_n$ symmetries, where $n$
is an even number. All realistic models involve some extra shaping symmetry that can help to control the presence of operators in the sectors responsible for Yukawa couplings and vacuum alignment, so 
the idea of an extra shaping symmetry is not new. What is new is that 
our mechanism restricts the shaping symmetries to be strings of discrete
symmetries such as $\mathbb{Z}_2$ and $\mathbb{Z}_4$ symmetries in order that
the vevs of the flavon fields be forced to be purely real or purely
imaginary. In particular, this prohibits the use of continuous shaping
symmetries such as for example a $U(1)$ symmetry; discrete symmetries like,
e.g.\ $\mathbb{Z}_3$ or $\mathbb{Z}_5$  symmetries are also not suitable as
they would not lead to purely real or purely imaginary vevs. 
This means that many of the existing models which have been proposed to describe tri-bimaximal lepton mixing are not viable for explaining the right-angled unitarity triangle, and we are forced to invent new models.

(iv) {\bf Spontaneous CP violation}:
 Another requirement of our mechanism is that CP is conserved in the theory at the high energy scale,
and is only broken spontaneously by the (complex) vevs of flavons. Such a scenario has been proposed 
previously in order to account for the smallness of CP violation in the soft SUSY sector~\cite{spontCPV}. Here it will be an essential
ingredient in obtaining the prediction of $\alpha \approx 90^\circ$. Thus we envisage 
models with family symmetries and spontaneous CP violation, in which the flavour structure as well as CP violation are generated from family symmetry breaking. 

In section~2 we explain our mechanism in general terms. In sections~3 and~4
we then turn to two realistic examples of 
$SU(5)$ GUT models with $A_4$ and $S_4$ family symmetries, respectively, plus extra  $\mathbb{Z}_n$ shaping symmetries. The $A_4\times SU(5)$ model in section~3 is an example of an indirect 
model, similar in nature to the model proposed in \cite{Antusch:2010es}, 
while the $S_4\times SU(5)$ model in section~4 is an example of a direct 
model, similar in nature to the model proposed in \cite{Hagedorn:2010th}.
The models are modified here to take account of the restriction on the shaping symmetry in (iii).
Both models illustrate that the right-angled CKM unitarity triangle can indeed be understood from an underlying flavour model with discrete symmetries which were introduced previously for the purpose of providing as an explanation of tri-bimaximal lepton mixing. In both models 
the quark mass matrices have 1-3 texture zeros and
satisfy the ``phase sum rule'' derived in \cite{Antusch:2009hq},
while the lepton mass matrices lead to the ``lepton mixing sum rule'' 
\cite{sumrule1,sumrule2}
together with a new prediction that the 
leptonic CP violating oscillation phase is close to either $0^\circ$, $90^\circ$,
$180^\circ$, or $270^\circ$ depending on the model,
with neutrino masses being purely real (no complex Majorana phases).
We summarise and conclude the paper in section~5, providing appendices on the 
more technical aspects of the models including their ultraviolet completion.

\section{The real/imaginary vacuum alignment mechanism}
\label{sec:general}
The goal of this paper is to show how the nearly right-angled CKM unitarity triangle (i.e.\ $\alpha \approx 90^\circ$) can be explained in flavour models. In this section we describe in general terms how this can be achieved in theories with spontaneous CP violation via purely real or imaginary alignments for the flavon fields. The flavon fields break the family symmetry, give rise to the flavour structure, and have to induce the observed CP violation via their vevs. 

\subsection{Motivation: The phase sum rule}
The motivation for this approach is provided by the phase sum rule of \cite{Antusch:2009hq},
which states that if the 1-3 mixing in both, the up-type quark mass matrix as well as the down-type mass matrix vanish (approximately), then there holds the following relation for the angle $\alpha$ of the CKM unitarity triangle:
\begin{equation}
\alpha \approx \delta_{12}^d - \delta_{12}^u \;.
\end{equation} 
The phases $\delta_{12}^d$ and $\delta_{12}^u$ are the arguments of the complex 1-2 rotation angles in the up-type and down-type quark mass matrices, as defined in  \cite{Antusch:2009hq}.

To explain $\alpha \approx 90^\circ$ one might therefore simply try to realise $\delta_{12}^d = 90^\circ$, $\delta_{12}^u = 0$ or alternatively $\delta_{12}^d = 0$, $\delta_{12}^u = -90^\circ$ in a model of flavour. For hierarchical mass matrices, this corresponds to the 1-2 and the 2-2 elements of the mass matrix being either purely real or imaginary. When the Yukawa matrices are generated after the breaking of some (non-Abelian) family symmetry, we thus need vevs of the flavons which have either purely real or purely imaginary components.

We now discuss in general terms how this might be achieved in scenarios with discrete symmetries in addition to non-Abelian family symmetries and later on we will also give two concrete examples.

\subsection{Method: Discrete vacuum alignment}
As mentioned above, we assume spontaneous CP violation, i.e.\ that CP violation is induced via the vevs of the flavons only, whereas the fundamental theory conserves CP. More specifically, we will assume that in the phase of unbroken family symmetry, there exists a basis where all parameters are real. 

Furthermore, we will consider the case that in addition to a non-Abelian discrete family symmetry, the flavour model features extra 
$\mathbb{Z}_n$ shaping symmetries. When a flavon $\phi$ carries single charge under the $\mathbb{Z}_n$ symmetry ($n\ge2$), then typical terms in the flavon superpotential, which ``drive'' the flavon vev non-zero, have the form
\begin{equation}\label{eq:flavonpotentialZn}
P \left( \frac{\phi^n}{\Lambda^{n-2}} \mp M^2 \right).
\end{equation} 
The field $P$ is the so-called ``driving superfield'', meaning that the $F$-term $|F_P|^2$ generates the potential for the scalar component of $\phi$ which enforces a non-zero vev. Here and in the following we will use the same letters for the superfields and its component fields. $\Lambda$ is the (real and positive) suppression scale of the effective operator, typically associated with the mass(es) of the messenger field(s) involved in its generation, and $M$ here is simply a (real) mass scale. 
From the potential for $\phi$,
\begin{equation}
|F_P|^2 =  \left| \frac{\phi^n}{\Lambda^{n-2}} \mp M^2 \right|^2 ,
\end{equation} 
we see that the vev of $\phi$ has to satisfy
\begin{equation}\label{eq:flavonvevsdiscrete}
\phi^n = \pm \,\Lambda^{n-2}  M^2\;.
\end{equation} 

The final step to explain our method is to argue that, whenever the flavon vev depends on just one single parameter, Eq.~\eqref{eq:flavonpotentialZn} forces the phase of the flavon vev to take only certain discrete values. For instance, in the simplest case where $\phi$ is a singlet under the non-Abelian family symmetry, it is clear that the phase is determined to be:
\begin{equation}\label{eq:phaseswithZn}
\arg(\langle \phi \rangle) = \left\{ \begin{array}{ll}
\frac{2 \pi}{n}q \;,\quad q = 1, \dots , n & \mbox{\vphantom{$\frac{f}{f}$} for ``$-$'' in Eq.~(\ref{eq:flavonpotentialZn}),}\\
\frac{2 \pi}{n} q +\frac{\pi}{n} \;,\quad q = 1, \dots , n & \mbox{\vphantom{$\frac{f}{f}$} for ``$+$'' in Eq.~(\ref{eq:flavonpotentialZn}).}
\end{array}
\right.
\end{equation} 

For example with a $\mathbb{Z}_2$ symmetry and a ``$-$''-sign in Eq.~(\ref{eq:flavonpotentialZn}) the phase satisfies $\arg(\langle \phi \rangle) \in \{0,\pi\}$ and thus the vev is real. For the ``$+$''-sign we have $\arg(\langle \phi \rangle) \in \{\pi/2,3\pi/2\}$ and the vev is purely imaginary.
Similarly, with a $\mathbb{Z}_4$ symmetry, we see that for the ``$-$''-sign the phase can take the possible values $\arg(\langle \phi \rangle) \in \{0,\pi/2,\pi,3\pi/2\}$ and for the ``$+$''-sign it can take the values $\arg(\langle \phi \rangle) \in \{\pi/4, 3 \pi/4, 5 \pi/4, 7 \pi/4\}$. So only for the ``$-$''-sign the flavon vev is either purely real or purely imaginary.\footnote{We note that with $\mathbb{Z}_n$ symmetries other than $\mathbb{Z}_2$ or $\mathbb{Z}_4$ one obtains different discrete possibilities for the phases which may also be interesting for model building. In this paper, however, we will focus on $\mathbb{Z}_2$ and $\mathbb{Z}_4$ symmetries since we are interested in purely real or purely imaginary flavon vevs.}
As we argued in the previous subsection, such either purely real or purely imaginary aligned flavons will be the building blocks for the proposed explanation of the right-angled CKM unitarity triangle.

The above arguments continue to hold true if the flavons are, for example, triplets of the non-Abelian family symmetry. In fact, typically, in explicit models we will deal with flavons which (by means of other terms in the flavon potential) are forced to point in specific directions in flavour space and thus depend only on one continuous parameter, say $x$. Examples for such flavons may be 
\begin{equation}
\phi_3 \propto \begin{pmatrix} 0 \\ 0 \\ x \end{pmatrix},\;
\phi_{23} \propto \begin{pmatrix} 0 \\ x \\ -x \end{pmatrix}\; 
\mbox{or}\;\;  \phi_{123} \propto \begin{pmatrix} x \\ x \\ x \end{pmatrix} .
\end{equation}
When the vevs of such flavons are driven by terms as in Eq.~(\ref{eq:flavonpotentialZn}), the phases are again forced to take only values as specified in Eq.~(\ref{eq:phaseswithZn}).

We have argued in this section that flavons which have either purely real or purely imaginary vevs could be the building blocks for flavour models capable of explaining the nearly right-angled CKM unitarity triangle (i.e.\ $\alpha \approx 90^\circ$). A model-independent discussion and a derivation of the phase sum rule can be found in \cite{Antusch:2009hq}. We have outlined a possible method of how such purely real or purely imaginary flavons can be realised in models. The next step will be to apply the method to construct two example models featuring $\alpha \approx 90^\circ$.

\section{$\boldsymbol{A_4\times SU(5)}$}
As a first example we will now discuss a model based on an $SU(5)$ GUT with $A_4$ family symmetry (plus extra discrete symmetries and an $R$-symmetry), broken by the vevs of five flavon fields $\phi_1, \phi_2, \phi_3, \phi_{23}$ and $\phi_{123}$. This may be regarded as a variation of the 
indirect $A_4\times SU(5)$ model in \cite{Antusch:2010es}.
We start with the discussion of the flavon potential. Following the method described in the previous section, we use as discrete symmetries only $\mathbb{Z}_2$ and $\mathbb{Z}_4$ symmetries, such that the vevs will be either purely real or imaginary. The field content and the symmetries are listed in Tabs.\ 
\ref{tab:A4FlavonSector} and \ref{tab:A4MatterSector} for the flavon sector and the matter sector, respectively.  The complete messenger sector of the model will be presented in App.\ \ref{App:A4Messenger}.

\subsection{Flavon sector}

\begin{table}
\centering
\begin{tabular}{cccccccccc} \toprule 
& $SU(5)$ & $A_4$   & $\mathbb{Z}^{(1)}_4$ & $\mathbb{Z}^{(2)}_4$ & $\mathbb{Z}^{(3)}_4$ & $\mathbb{Z}^{(4)}_4$ & $\mathbb{Z}^{(1)}_2$ & $\mathbb{Z}_2^{(2)}$ & $U(1)_R$ \\ \midrule
\multicolumn{9}{l}{Flavons} \\ \midrule
$\phi_1$     & $\mathbf{1}$ & $\mathbf{3}$ & 3 & 0 & 0 & 0 & 0 & 0 & 0 \\
$\phi_2$     & $\mathbf{1}$ & $\mathbf{3}$ & 3 & 3 & 0 & 0 & 0 & 0 & 0 \\
$\phi_3$     & $\mathbf{1}$ & $\mathbf{3}$ & 0 & 0 & 0 & 0 & 1 & 0 & 0 \\
$\phi_{123}$ & $\mathbf{1}$ & $\mathbf{3}$ & 0 & 0 & 3 & 0 & 0 & 0 & 0 \\
$\phi_{23}$  & $\mathbf{1}$ & $\mathbf{3}$ & 0 & 0 & 3 & 3 & 0 & 0 & 0 \\ 
$\xi$ & $\mathbf{1}$ & $\mathbf{1}$ & 0 & 0 & 1 & 0 & 0 & 0 & 0 \\
\midrule
\multicolumn{9}{l}{Driving Fields} \\ \midrule
$P_i$ & $\mathbf{1}$ & $\mathbf{1}$ & 0 & 0 & 0 & 0 & 0 & 0 & 2 \\
$A_1$ & $\mathbf{1}$ & $\mathbf{3}$ & 2 & 0 & 0 & 0 & 0 & 0 & 2 \\
$A_2$ & $\mathbf{1}$ & $\mathbf{3}$ & 2 & 2 & 0 & 0 & 0 & 0 & 2 \\
$A_3$ & $\mathbf{1}$ & $\mathbf{3}$ & 0 & 0 & 0 & 0 & 0 & 0 & 2 \\
$A_{123}$ & $\mathbf{1}$ & $\mathbf{3}$ & 0 & 0 & 2 & 0 & 0 & 0 & 2 \\ 
$O_{1;2}$ & $\mathbf{1}$ & $\mathbf{1}$ & 2 & 1 & 0 & 0 & 0 & 0 & 2 \\
$O_{1;3}$ & $\mathbf{1}$ & $\mathbf{1}$ & 1 & 0 & 0 & 0 & 1 & 0 & 2 \\
$O_{2;3}$ & $\mathbf{1}$ & $\mathbf{1}$ & 1 & 1 & 0 & 0 & 1 & 0 & 2 \\ 
$O_{1;23}$ & $\mathbf{1}$ & $\mathbf{1}$ & 1 & 0 & 1 & 1 & 0 & 0 & 2 \\
$O_{123;23}$ & $\mathbf{1}$ & $\mathbf{1}$ & 0 & 0 & 2 & 1 & 0 & 0 & 2 \\ 
 \bottomrule
\end{tabular}
\caption{Flavon and driving fields of the $A_4$ model. Note that depending on which option for the alignment of $\phi_{123}$ is chosen, the fields $\xi$ and $A_{123}$ are present or not, as described in the main text. The $\mathbb{Z}_2^{(2)}$ symmetry is not necessary for the alignment itself, but it will be required for the matter (cf.\ Tab.\ \ref{tab:A4MatterSector}) and the messenger sectors (cf.\ App.\ \ref{App:A4Messenger}).  \label{tab:A4FlavonSector}
} 
\end{table}

In Sec.~\ref{sec:general} we have discussed in general terms how the phases of flavon vevs may be predicted from the flavon potential. In this section we will apply this method to construct a simple $A_4$ model capable of predicting a CKM unitarity triangle with $\alpha \approx 90^\circ$. As we will see, the alignment we produce here enables us to predict the CKM phase correctly and also give predictions for all the phases in the lepton sector which are not yet experimentally determined. The flavons and driving fields and their charges under the imposed symmetries are given in Tab.\ \ref{tab:A4FlavonSector}.

In addition to the flavons $\phi_1, \phi_2, \phi_3, \phi_{23}$ and $\phi_{123}$ responsible for the flavour structure, the table also contains the ``auxiliary flavon'' $\xi$, which will be used to align the vev of $\phi_{123}$. 
The ``driving fields'' will be called $P_i$, $A_i$ and $O_{i;j}$, and we use a
notation that via their $F$-term contributions to the flavon potential, the
$A_4$ singlet fields $P_i$ fix the phase as discussed in
Sec.~\ref{sec:general}, the triplets $A_i$ force flavons to point into certain
specific directions in flavour space, and the singlet fields $O_{i;j}$ align
the vev of the flavon $\phi_i$ to be orthogonal to the one of $\phi_j$. 
Notice that the driving fields $A_i$ and $O_{i;j}$ are all distinguished from
one another by their charge assignments, whereas the driving fields
$P_i$ are completely neutral under all shaping symmetries. Hence, in a generic
basis, each of these $P_i$ fields couples to the same set of terms with
different coupling constants. In order to apply the method outlined in
Sec.~\ref{sec:general} it is necessary to disentangle the equations by a
suitable basis transformation, the details of which are presented in
App.~\ref{app-basis}. 
For the sake of simplicity we will sometimes suppress (real and positive) order one coupling constants where they are not relevant for the model predictions.
In dealing with $A_4$ we will use the standard ``$SO(3)$ basis'' for which the singlet of ${\bf 3 \otimes 3}$ is given by the $SO(3)$-type inner product. The two triplets of ${\bf 3 \otimes 3}$ are constructed from the usual (antisymmetric) cross product '$\times$' and the symmetric star product '$\star$' (see, for example, \cite{King:2006np}).
The symmetric product is defined analogous to the cross product but with a relative plus sign instead of a minus sign.

Let us now start our discussion of the alignment by specifying the required form of the vevs of $\phi_1$, $\phi_2$, $\phi_3$, $\phi_{23}$,  $\phi_{123}$ for the construction of the $A_4$ model with $\alpha = 90^\circ$: 
\begin{equation} \label{eq:A4Alignment}
\langle \phi_1 \rangle \propto \begin{pmatrix} 1 \\ 0 \\ 0 \end{pmatrix} , \; \langle \phi_2 \rangle \propto \begin{pmatrix} 0 \\ -\mathrm{i} \\ 0 \end{pmatrix} ,\; \langle \phi_3 \rangle \propto \begin{pmatrix} 0 \\ 0 \\ 1 \end{pmatrix} , \;
\langle \phi_{23} \rangle \propto \begin{pmatrix} 0 \\ 1 \\ -1 \end{pmatrix} , \; \langle \phi_{123} \rangle \propto \begin{pmatrix} 1 \\ 1 \\ 1 \end{pmatrix}.
\end{equation}
For predicting $\alpha \approx 90^\circ$, one possibility will be to have an imaginary $\phi_2$ and real $\phi_1$, $\phi_3$, $\phi_{23}$, $\phi_{123}$. This is explicitly incorporated in Eq.\ \eqref{eq:A4Alignment} by assuming real proportionality constants here and in the following. The vevs of $\phi_{23}, \phi_{123}$ are familiar from various flavour models and lead to tri-bimaximal neutrino mixing. 

For realising the required vacuum alignment of $\phi_{123}$, we discuss two options: 

\begin{itemize}
\item {\bf Option A:} The (super-)potential for the first option is 
\begin{equation} \label{eq:A4Phi123AlignmentA}
\begin{split}
W_{\phi_{123}} =&  \; P_{123} \left( \frac{\phi_{123}^4}{M_{\Upsilon_{123;123}}^2} - \lambda \frac{ (\phi_{123} \phi_{123})_{1'} (\phi_{123} \phi_{123})_{1''} }{M_{\Upsilon'} M_{\Upsilon''} }  - M_{123}^2 \right) \;, \\
V_{\mathrm{soft}} =& \; m_{123}^2 |\phi_{123}|^2 \;,
\end{split}
\end{equation}
where $V_{\mathrm{soft}}$ displays a SUSY-breaking soft mass term for
$\phi_{123}$ with positive $m_{123}^2$. The brackets $(...)_{1'}$ and $(...)_{1''}$ mean that the fields are contracted to ${\bf{1'}}$ and ${\bf{1''}}$ representations of $A_4$. $\phi_{123}^4$ stands for 
$ (\phi_{123} \phi_{123})_{1} (\phi_{123} \phi_{123})_{1}$. The (real) constants $M_{\Upsilon_{123;123}}$, $M_{\Upsilon'}$, and $M_{\Upsilon''}$ denote messenger masses, see App.~\ref{App:A4Messenger}.

It is not obvious, that this potential gives the desired alignment of $\langle \phi_{123} \rangle \propto (\pm 1, \pm 1,\pm 1)$ or $(\pm i, \pm i,\pm i)$. Let us assume that $\langle \phi_{123}\rangle=  (x,y,z)$, then the invariant $(\ldots)_{1'} (\ldots)_{1''}$ gives a contribution to the $F$-term conditions of the form
$$
x^4 + y^4 + z^4 - x^2 y^2 - x^2 z^2 - y^2 z^2 \;.
$$
This combination obviously vanishes for $x^2 = y^2 = z^2$, which is already the desired alignment. Nevertheless, having only this invariant coupling to a driving field is not sufficient since the scale of $\langle \phi_{123} \rangle$ is completely arbitrary up to this stage. This is fixed by the $\phi_{123}^4$ term and the soft mass. Indeed, we have checked numerically that for $0 < \lambda < M_{\Upsilon'} M_{\Upsilon''} / M_{\Upsilon_{123;123}}^2$ we end up in a vacuum, where  $\langle \phi_{123} \rangle \propto (\pm 1, \pm 1,\pm 1)$ or $(\pm i, \pm i,\pm i)$, if we assume $M_{123}$ to be real.

\item {\bf Option B:} Alternatively one can achieve the alignment of the vev of $\phi_{123}$ by: 
\begin{equation}  \label{eq:A4Phi123AlignmentB}
W'_{\phi_{123}} =  A_{123} (\xi \phi_{123} + \phi_{123} \star \phi_{123})  + P_{123} \left( \frac{\phi_{123}^4}{M_{\Upsilon_{123;123}}^2} +  \frac{\xi^2 \phi_{123}^2}{M_{\Upsilon_{123;123}}^2}  + \frac{\xi^4}{M_{\Upsilon_{123;123}}^2}  - M_{123}^2  \right)   \;,
\end{equation}
The $F$-term equations for $A_{123}$ and $P_{123}$ have three distinct solutions.
With $\langle \xi \rangle = 0$ the potential reduces to the case in Eq.\ \eqref{eq:A4PhiAlignment} giving, e.g.\ $ \langle \phi_{123} \rangle \propto (1,0,0)$. For $\langle \xi \rangle \neq 0$ there exist two possibilities for $\langle \phi_{123} \rangle$: the vev of $\phi_{123}$ could vanish or, alternatively, point into the directions $\langle \phi_{123} \rangle \propto (\pm 1, \pm 1,\pm 1)$. The latter can be understood in the following way. The $F$-term conditions of the first term in Eq.\ \eqref{eq:A4Phi123AlignmentB} give three relations between the $\xi$ vev and the components of $\langle\phi_{123}\rangle$ which enter the equations cyclically preferring already the solution $(\pm 1, \pm 1,\pm 1)$. We only need a term, which drives the vevs to non-zero values which is done by the second term in Eq.\ \eqref{eq:A4Phi123AlignmentB}.

In the following we assume, that the latter option is realised.
Compared to option A, the  ``auxiliary flavon'' $\xi$ and the additional driving field $A_{123}$ are introduced, however no soft terms are involved in the alignment.
\end{itemize}

Now that we have this alignment at hand the alignment of the other flavons is comparatively straightforward. 
The vevs of the other flavons are determined by the following additional superpotential terms: 
\begin{align}
W_{\phi_1,\phi_2,\phi_3} =& \; P_1  \left( \frac{\phi_1^4}{M_{\Upsilon_{1;1}}^2}  - M_1^2 \right) + P_2  \left( \frac{\phi_2^4}{M_{\Upsilon_{2;2}}^2}  - M_2^2 \right) + P_3 (\phi_3^2 - M_3^2) \nonumber\\ 
&+ A_i (\phi_i \star \phi_i) + O_{ij} ( \phi_i . \phi_j )\;,  \label{eq:A4PhiAlignment} \\ 
W_{\phi_{23}} =&  \; P_{23} \left( \frac{\phi_{23}^4}{M_{\Upsilon_{23;23}}^2} - M_{23}^2 \right)  + O_{1;23} (\phi_1 . \phi_{23}) + O_{123;23} (\phi_{123} . \phi_{23})\;. \label{eq:A4Phi23Alignment}
\end{align}
Note that in Eq.\ \eqref{eq:A4PhiAlignment} only the symmetric coupling $A_i (\phi_i \star \phi_i)$ appears, since the cross product vanishes for two identical fields (which is of course also allowed by the symmetries). As discussed in Sec.~\ref{sec:general}, and assuming spontaneous CP violation, we obtain that the phases of the vevs of $\phi_1$,  $\phi_2$, $\phi_3$, $\phi_{23}$, $\phi_{123}$ can only take the values $\{0,\pi, \pm \pi/2\}$, as desired. Among these possible vacua, we will concentrate in the following on the solution where $\phi_2$ is purely imaginary and the other flavon vevs are real.\footnote{We note that there are other combinations of the flavon vevs'  phases leading to the same results, whereas others are phenomenologically invalid. In principle, higher-dimensional (Planck scale suppressed) operators may violate the discrete symmetries and favour one vacuum over the others. This preferred vacuum may then expand and finally be the only one in our observable universe. 
In a more fundamental theory one may even attempt to calculate which vacua are preferred, but for the present work this is clearly beyond the scope. 
}

Together, the flavon superpotential in Eqs.\ \eqref{eq:A4Phi123AlignmentA}-\eqref{eq:A4Phi23Alignment} can result in the desired flavon alignment with imaginary $\phi_2$ and real $\phi_1,\phi_3,\phi_{23}, \phi_{123}$ as specified in Eq.~(\ref{eq:A4Alignment}).

\subsection{Matter sector and predictions}

\begin{table}
\centering
\begin{tabular}{cccccccccc} \toprule 
& $SU(5)$ & $A_4$   & $\mathbb{Z}^{(1)}_4$ & $\mathbb{Z}^{(2)}_4$ & $\mathbb{Z}^{(3)}_4$ & $\mathbb{Z}^{(4)}_4$ & $\mathbb{Z}^{(1)}_2$ & $\mathbb{Z}_2^{(2)}$ & $U(1)_R$ \\ \midrule
\multicolumn{9}{l}{Matter Fields} \\ \midrule
$F$   & $\mathbf{\overline{5}}$  & $\mathbf{3}$ & 0 & 0 & 0 & 0 & 0 & 1 & 1 \\ 
$T_1$ & $\mathbf{10}$ & $\mathbf{1}$ & 1 & 0 & 0 & 0 & 0 & 0 & 1 \\
$T_2$ & $\mathbf{10}$ & $\mathbf{1}$ & 0 & 0 & 1 & 0 & 0 & 0 & 1 \\
$T_3$ & $\mathbf{10}$ & $\mathbf{1}$ & 0 & 0 & 0 & 0 & 1 & 0 & 1 \\ 
$N_1$ & $\mathbf{1}$ & $\mathbf{1}$ & 0 & 0 & 3 & 3 & 0 & 0 & 1 \\
$N_2$ & $\mathbf{1}$ & $\mathbf{1}$ & 0 & 0 & 3 & 0 & 0 & 0 & 1 \\ \midrule
\multicolumn{9}{l}{Higgs Fields} \\ \midrule
$\bar{H}_1$ & $\mathbf{\overline{5}}$ & $\mathbf{1}$ & 0 & 1 & 0 & 0 & 0 & 0 & 0 \\
$\bar{H}_2$ & $\mathbf{\overline{45}}$ & $\mathbf{1}$ & 0 & 0 & 0 & 1 & 0 & 0 & 0 \\
$\bar{H}_3$ & $\mathbf{\overline{5}}$ & $\mathbf{1}$ & 0 & 0 & 0 & 0 & 0 & 0 & 0 \\
$H$               & $\mathbf{5}$ & $\mathbf{1}$ & 0 & 0 & 0 & 0 & 0 & 0 & 0 \\ 
$H_{24}$     & $\mathbf{24}$ & $\mathbf{1}$ & 0 & 0 & 0 & 0 & 0 & 1 & 0 \\   \bottomrule
\end{tabular}
\caption{The matter and Higgs fields of the $A_4$ model. \label{tab:A4MatterSector}} 
\end{table}

With the $A_4$ breaking flavon sector and the alignment of the flavon vevs at hand we will now turn to the fermion masses and mixings within the $A_4 \times SU(5)$ GUT model. The matter content of the Standard Model fits into the five-dimensional representation of $SU(5)$
\begin{equation}
F_i =  \begin{pmatrix}
      d_R^{c} & d_B^{c} & d_G^{c} & e &-\nu
       \end{pmatrix}_i \:,
\end{equation}
which we assume to be a triplet under $A_4$, and the ten-dimensional representations of $SU(5)$
\begin{equation}
T_i = \frac{1}{\sqrt{2}}
        \begin{pmatrix}
        0 & -u_G^{c} & u_B^{c} & -u_{R} & -d_{R} \\
        u_G^{c} & 0 & -u_R^{c} & -u_{B} & -d_{B} \\
        -u_B^{c} & u_R^{c} & 0 & -u_{G} & -d_{G} \\
        u_{R} & u_{B} & u_{G} & 0 & -e^c \\
        d_{R} & d_{B} & d_{G} & e^c & 0
        \end{pmatrix}_i \:,
\end{equation}
which we assume to be singlets under $A_4$. We also add two right-handed neutrinos $N_1$ and $N_2$, being singlets under $SU(5)$ and $A_4$, to generate masses for two of the light neutrinos via the seesaw mechanism \cite{seesaw}.
The Higgs sector consists of $H_{24}$, $H$ and $\bar{H}_i$, $i = 1,2,3$. $H_{24}$ is the GUT symmetry breaking Higgs field while $H$ contains the MSSM up-type Higgs doublet, and the down-type Higgs doublet is a linear combination of the doublet components of the $\bar{H}_i$ fields.

The model will predict the GUT scale ratios $y_\tau/y_b$ and $y_\mu/y_s$.  Instead of the commonly encountered $b$-$\tau$ Yukawa unification and the Georgi--Jarlskog relation \cite{Georgi:1979df} for $y_\mu/y_s$, which are phenomenologically somewhat challenged in CMSSM scenarios \cite{Altmannshofer:2008vr, Antusch:2008tf, Antusch:2009gu}, our model predicts the GUT scale relations \cite{Antusch:2009gu}
\begin{equation}
\frac{y_\mu}{y_s} \approx \frac{9}{2} \quad \text{and} \quad \frac{y_\tau}{y_b} = -\frac{3}{2} \;,
\end{equation}
which differ from the Georgi-Jarlskog predictions by an overall factor of 3/2 giving 
better agreement with phenomenology. The operators yielding these predictions contain $H_{24}$ with its vev given by the diagonal matrix
\begin{equation}
\langle H_{24} \rangle \sim v_{24} \:\text{diag} (1,1,1-3/2,-3/2) \;.
\end{equation} 

The non-renormalisable superpotential which is generated after integrating out the messenger fields, cf.\ App.\ \ref{App:A4Messenger}, is given by the following terms:
\begin{align} 
W_d =& \; F H_{24} \left(  \frac{ T_1 \bar{H}_1 \phi_2 }{M_{\Xi_2} M_{\Xi'_2}} + \frac{ T_2 \bar{H}_3 \phi_{123} }{M_{\Xi_{123}} M_{\Xi'_{123}}} + \frac{ T_2 \bar{H}_2 \phi_{23} }{M_{\Xi_{23}} M_{\Xi'_{23}}}  +   \frac{ T_3 \bar{H}_3 \phi_3 }{M_{\Xi_3} M_{\Xi'_3}}  \right)   \;, \label{eq:A4YukawaD} \\
W_u =&  \; T_3^2 H +  \frac{ T_2^2 H \phi_{123}^2 }{M_{\Theta_{2;2}} M_{\Upsilon_{123;123}} } + \frac{ T_1^2 H  \phi_1^2 }{M_{\Theta_{1;1}} M_{\Upsilon_{1;1}} } +  \frac{ T_2 T_3 H \phi_{123} \phi_3 }{M_{\Theta_{2;3}} M_{\Upsilon_{3;123}} } \nonumber\\
 &  + \frac{T_1 T_3 H \phi_1 \phi_3}{M_{\Theta_{1;3}} M_{\Upsilon_{1;3}} } +   \frac{T_1 T_2 H \phi_{123} \phi_1}{M_{\Theta_{1;2}} M_{\Upsilon_{1;123}} } \;, \label{eq:A4YukawaU}\\
W_\nu =&  \; F H_{24} H \left( \frac{N_1 \phi_{23}}{M_{\Xi_{23}} M_{\Xi'_{23}}} + \frac{N_2 \phi_{123}}{M_{\Xi_{123}} M_{\Xi'_{123}}}  \right) \;, \label{eq:A4YukawaN}\\
W_N =&  \;  \frac{\phi_{23}^2 N^2_1}{M_{\Upsilon_{23}}} + \frac{\phi^2_{123} N_2^2}{M_{\Upsilon_{123}}}  \;. \label{eq:A4MN}
\end{align}
As before, order one coefficients are dropped where they have no influence on the model predictions. The new masses are the masses of the messenger fields, as will be discussed in App.~\ref{App:A4Messenger}.

For the low energy charged lepton and down-type quark Yukawa couplings we define
\begin{gather}
\epsilon_2 \sim \frac{v_{24} | \langle \phi_2 \rangle |}{M_{\Xi_2} M_{\Xi'_2}}  \;, \;\;
\epsilon_3 \sim \frac{v_{24} | \langle \phi_3 \rangle |}{M_{\Xi_3} M_{\Xi'_3}}  \;, \;\;
\epsilon_{23} \sim \frac{v_{24} | \langle \phi_{23} \rangle |}{M_{\Xi_{23} } M_{\Xi'_{23}}}  \;, \;\;
\epsilon_{123} \sim \frac{v_{24} | \langle \phi_{123} \rangle |}{M_{\Xi_{123} } M_{\Xi'_{123}}}  \;, 
\end{gather}
where we have dropped order one coefficients and Higgs mixing angles. Similarly for the up-type quark sector we define 
\begin{gather}
a_{11} \sim \frac{ |\langle \phi_1 \rangle|^2 }{ M_{\Theta_{1;1}} M_{\Upsilon_{1;1}} } \;,\;\;
a_{22} \sim \frac{ |\langle \phi_{123} \rangle|^2 }{ M_{\Theta_{2;2}} M_{\Upsilon_{123;123}} } \;,\;\;
a_{12} \sim \frac{ |\langle \phi_1 \rangle| |\langle \phi_{123} \rangle| }{ M_{\Theta_{1;2}} M_{\Upsilon_{1;123}} } \;,\;\;
a_{23} \sim \frac{ |\langle \phi_3 \rangle| |\langle \phi_{123} \rangle| }{ M_{\Theta_{2;3}} M_{\Upsilon_{3;123}} } \;.\;\;
\end{gather}
The top Yukawa coupling $y_t = a_{33}$ is generated at tree-level and the would-be $a_{13}$ vanishes due to the orthogonality of $\phi_1$ and $\phi_3$. For the neutrino Yukawa couplings we define
\begin{equation}
a_{\nu_1} \sim \frac{v_{24} |\langle \phi_{23} \rangle |}{M_{\Xi_{23}} M_{\Xi'_{23}}}   \quad \text{and} \quad a_{\nu_2} \sim \frac{v_{24} |\langle \phi_{123} \rangle |}{M_{\Xi_{123}} M_{\Xi'_{123}}}  \;,
\end{equation}
and for the right-handed neutrino masses
\begin{equation}
M_{R_1} \sim \frac{| \langle \phi_{23} \rangle|^2}{M_{\Upsilon_{23}}} \quad \text{and} \quad M_{R_2} \sim \frac{|\langle \phi_{123} \rangle|^2}{M_{\Upsilon_{123}}} \;.
\end{equation}

With these definitions at hand we can express the Yukawa couplings in a simple
form using the PDG convention \cite{PDG}, namely
\begin{equation}
\mathcal{L}_{\mathrm{Yuk}} = - (Y^*_d)_{ij} Q_i \bar{d}_j H_d - (Y^*_e)_{ij} L_i \bar{e}_j H_d  - (Y_u^*)_{ij} Q_i \bar{u}_j H_u + \mathrm{H.c.}\;.
\end{equation}
Regarding the quark Yukawa matrices, from Eqs.\ \eqref{eq:A4YukawaD} and \eqref{eq:A4YukawaU} and using the above definitions as well as the alignments of Eq.\ \eqref{eq:A4Alignment}  we obtain:
\begin{align}
Y_d &= \begin{pmatrix} 0 & \mathrm{i} \, \epsilon_2 & 0 \\ \epsilon_{123} & \epsilon_{23} + \epsilon_{123} & -\epsilon_{23} + \epsilon_{123} \\ 0 & 0 & \epsilon_3  \end{pmatrix}  \;, \\
Y_u &= \begin{pmatrix} a_{11} & a_{12} & 0 \\ a_{12} & a_{22} & a_{23} \\ 0 & a_{23} & a_{33} \end{pmatrix} \;.
\end{align}
Note that due to the complex conjugation in the definition of the Yukawa
couplings a factor of $+\text{i}$ appears now in the 1-2 element of $Y_d$
(instead of $-\text{i}$). We see that the ``phase sum rule'' of
Ref.~\cite{Antusch:2009hq} applies since in both the up and the down quark
sector we have zero 1-3 mixing.
As discussed in \cite{Antusch:2009hq}, the structures of $Y_u$ and $Y_d$ give
the correct quark masses and mixings including a CKM
matrix that features a right-angled unitarity triangle with $\alpha\approx
\pm 90^\circ$. In order to obtain the positive sign of $\alpha$ we
need to require a relative sign difference between the omitted real order one
coefficients of the 1-2 and 2-2 elements of either $Y_u$ or $Y_d$ (but not
both).
Note that the moduli of the parameters $a_{ij}$ are not predicted in our model, since these Yukawa couplings stem from effective operators generated by messenger fields with (in general) different masses. They will be fixed by the fit to the up-type quark masses and the quark mixing angles. 

For the neutrino and charged lepton sector we obtain:
\begin{align}
M_R &= \begin{pmatrix} M_{R_1} & 0 \\ 0 & M_{R_2} \end{pmatrix} \;, \\
Y_{\nu} &= \begin{pmatrix} 0 & a_{\nu_2} \\ a_{\nu_1} & a_{\nu_2} \\ - a_{\nu_1} & a_{\nu_2} \end{pmatrix} \;, \\
Y_e^T &= -\frac{3}{2} \begin{pmatrix} 0 & \mathrm{i} \, \epsilon_2 & 0 \\
 \epsilon_{123} & -3 \, \epsilon_{23} + \epsilon_{123} & 3 \, \epsilon_{23} +
 \epsilon_{123} \\ 0 & 0 & \epsilon_3  \end{pmatrix}\ .
\end{align}
The size of the neutrino Yukawa couplings is given by the two parameters $a_{\nu_1}$ and  $a_{\nu_2}$, which are of the order $\epsilon_{23}$ and $\epsilon_{123}$. The right-handed neutrino masses $M_{R_1}$ and $M_{R_2}$ can be chosen such that the two observed neutrino mass squared differences are obtained, with one of the light neutrinos being massless (by construction since we have assumed only two right-handed neutrinos).

The mixing in the neutrino sector is ``tri-bimaximal'' to a good approximation, since the neutrino Yukawa matrix $Y_{\nu}$ above satisfies the conditions of constrained sequential dominance~\cite{sumrule1}. In the considered $SU(5)$ GUT framework, $Y_e$ is connected to $Y_d$ and we obtain predictions for the lepton mixing parameters due to ``charged lepton'' corrections.

The alignment of Eq.\ \eqref{eq:A4Alignment} fixes all the phases in the lepton sector as well, leading to a Maki-Nakagawa-Sakata (MNS) mixing matrix with $\delta_{\text{MNS}} \approx 0^\circ$ or  $180^\circ$, depending on the relative sign of $\epsilon_{123}$ and $\epsilon_{23}$, and vanishing CP violating Majorana phases.\footnote{We remark that there exist other similar vacuum alignment possibilities for $\alpha \approx 90^\circ$ leading to other discrete predictions for $\delta_{\text{MNS}}$, i.e.\ $\delta_{\text{MNS}} \in \{0,\pi/2,\pi,3\pi/2\}$. A scenario with maximal leptonic CP violation is discussed in Sec.~\ref{sec-s4}. Strictly speaking, our approach is in general only predicting one out of these discrete possibilities. With a specific alignment chosen, here with the one in Eq.\  \eqref{eq:A4Alignment}, the values of all phases (including also the two Majorana phases) can be calculated.} The would-be leptonic unitarity triangle thus collapses to a line. Combining tri-bimaximal neutrino mixing with the charged lepton corrections the predictions satisfy the lepton mixing sum rule $\theta^{\text{MNS}}_{12} - \cos(\delta_{\text{MNS}})\theta^{\text{MNS}}_{13} \approx \arcsin (1/\sqrt{3})$ \cite{sumrule1,sumrule2}. With $\theta^{\text{MNS}}_{13} \approx 3^\circ$, the $A_4$ model therefore predicts a $\pm 3^\circ$ shift of the solar mixing angle from its tri-bimaximal value of $35.26^\circ$.

In summary, we have constructed a simple model based on $A_4 \times SU(5)$ symmetry, plus discrete $\mathbb{Z}_2$ and $\mathbb{Z}_4$ shaping symmetries, which is capable of predicting a right-angled CKM unitarity triangle $\alpha \approx 90^\circ$ following the method of ``discrete vacuum alignment'' described in Sec.\ \ref{sec:general}.

\section{\label{sec-s4}$\boldsymbol{S_4\times SU(5)}$}

In this section we present a variation of the direct $S_4\times SU(5)$ model in
\cite{Hagedorn:2010th}. We adopt the same $S_4$ basis as well as the
same notation; for the sake of brevity we refer the reader to
\cite{Hagedorn:2010th} whenever appropriate. 
As the neutrino sector remains unaltered we do not delve into an in-depth
discussion thereof. 
With the quark sector being our primary focus we wish to accommodate the
right-angled CKM unitary triangle by means of real and imaginary entries in
the quark mass matrices. 
Thus we are led to drop the flavon field $\wt \phi^u_2$
and introduce two new ones, $\phi^u_{1'}$ and $\wt \phi^d_2$. This entails 
slight changes in the choice of superfields which drive the flavon vevs.
We begin by briefly stating the leading Yukawa superpotential terms, the
assumed vacuum configurations and the resulting quark mass matrices. We
proceed by discussing the required flavon potential in detail.
The $U(1)$ shaping symmetry which was introduced in \cite{Hagedorn:2010th} in
order to control the allowed terms must be replaced by a set of $\mathbb Z_n$
symmetries as discussed in Sec.~\ref{sec:general}.
We construct all possible such symmetries that can help constrain our
model at the effective level and determine all allowed terms which - if
present - would spoil the desired structure. This investigation shows that
even the most general set of allowed $\mathbb Z_n$ symmetries is insufficient
to yield a viable model. However, we argue that all additional
dangerous terms can be forbidden by invoking a set of  messenger fields which
gives rise to the required terms but disallows the dangerous ones.

\subsection{Outline of the model}

To make the following self-contained, we remark that the
${\bs{\overline{5}}}$ of $SU(5)$ is denoted by $F$ while the $\bs{10}$ is
written as $T$. They furnish the following  $S_4$ representations:
$$
F ~=~ \begin{pmatrix}F_1\\F_2\\F_3\end{pmatrix} ~\sim~ \bs{3} \ , \qquad 
T ~=~ \begin{pmatrix}T_1\\T_2\end{pmatrix} ~\sim~ \bs{2} \ , \quad
T_3 ~\sim~ \bs{1} \ .
$$
The desired Yukawa superpotential terms are
\bea
W_u&=& T_3 T_3 H_5 
+ \frac{1}{M} TT\phi^u_2 H_5
+ \frac{1}{M^2} TT (\phi^u_{1'})^2 H_5 
+\frac{1}{M^3} TT(\phi^d_3)^2\phi^\nu_1H_5\ ,\label{up1}\\
W_d&=&\frac{1}{M} FT_3 \phi^d_3 H_{\bar 5} 
+\frac{1}{M^2} (F\wt\phi^d_3)_1(T \phi^d_2)_1 H_{\overline{45}}  
+ \frac{1}{M^2} (F\phi^d_3)_2 (T \wt\phi^d_2)_2 H_{\bar 5} \
,\label{down1}\\[2.5mm] 
W_\nu&=&FNH_5+~
N(\phi^\nu_{3'} + \phi^\nu_{2}+\phi^\nu_{1})N  \ , \label{nu2}
\eea
where $(\cdots)_{1,2}$ denotes $S_4$ contractions to the ${\bf 1,2}$
representations, respectively. Here and in the following we assume all order one
coefficients to be real and suppress them in our notation.  Furthermore, 
the non-renormalisable terms are suppressed by a {\it common} mass scale~$M$. 

With the vacuum configuration of the flavon fields given as 
\be
\langle \phi^u_2 \rangle ~\sim~ \lambda^4 M \begin{pmatrix}0\\1\end{pmatrix},
\quad
\langle \phi^u_{1'} \rangle ~\sim~  \lambda^3 M \ ,\label{fup3}
\ee
\be
\langle \phi^d_3 \rangle ~\sim~ \lambda^{2} M \begin{pmatrix}0\\1\\0\end{pmatrix},
\quad  
\langle \wt\phi^d_3 \rangle ~\sim~ \lambda^{3}
M \begin{pmatrix}0\\2\\1\end{pmatrix}, 
\quad
\langle \phi^d_2 \rangle ~\sim~  \lambda M \begin{pmatrix}1\\0\end{pmatrix} ,
\quad
\langle \wt \phi^d_2 \rangle ~\sim~  \lambda^3 M \begin{pmatrix}-\mathrm{i}\\-\mathrm{i}\end{pmatrix},\label{fdown4}
\ee
\bea
\langle \phi^\nu_{1} \rangle ~\sim~ \lambda^4 M \ ,\qquad
\langle \phi^\nu_{2} \rangle ~\sim~ \lambda^4
M \begin{pmatrix}1\\1\end{pmatrix} , \qquad
\langle \phi^\nu_{3'} \rangle ~\sim~ \lambda^4
M \begin{pmatrix}1\\1\\1\end{pmatrix}  ,\label{fnu3}
\eea
we are led to the following quark mass matrices
\be
M_u \sim \begin{pmatrix}\lambda^8 &  \lambda^6 &0 \\
\lambda^6&\lambda^4&0\\0&0&1 \end{pmatrix} v_u \ ,\qquad
M_d \sim \begin{pmatrix}0 &  \mathrm{i}\lambda^{5} &0 \\
\mathrm{i}\lambda^{5}&\lambda^{4}&2\lambda^{4}\\0&0&\lambda^{2} \end{pmatrix}
v_d\ ,\label{mumd}
\ee
where $\lambda \approx 0.22$ denotes the sine of the Cabibbo angle.
In the up quark mass matrix, the 2-2 element arises from the second term
of Eq.~(\ref{up1}), the 1-2 and 2-1 elements originate from the third term,
and the 1-1 element from the fourth. As the vevs of all the flavon fields
occurring in Eq.~(\ref{up1})  are real, the matrix $M_u$ is real as well. 
Turning to the down quark mass matrix, we obtain the 3-3 entry from the first
term of Eq.~(\ref{down1}), while the 2-2 and 2-3 entries are derived from the
second term. Notice the relative factor of 2 in the 2-3 element which is
related to the alignment of $\wt\phi^d_3$ and serves to improve the fit of the
2-3 CKM mixing. From Eq.~(\ref{mumd}) we get $\theta_{23}^{\text{CKM}} \approx 2
\frac{m_s}{m_b}$; evaluating this at the scale of, e.g.\  the top mass yields a
value of around $0.038$ which is to be compared to the measured 2-3 CKM mixing
of $0.041$, see, e.g.\ \cite{Xing:2007fb}. Without the factor of
2 one would be far off. 
Finally the 1-2 and 2-1 elements of $M_d$ originate from the third term of
Eq.~(\ref{down1}). Due to the vev configurations of the down-type flavons of
Eq.~(\ref{fdown4}), we find purely imaginary entries for the 1-2 and 2-1
elements while the other elements of the down quark mass matrix are all real.
As discussed in \cite{Antusch:2009hq}, the structures of $M_u$ and $M_d$ of
Eq.~(\ref{mumd}), with zero 1-3 mixings and a non-vanishing contribution in
the 1-1 element of $M_u$, give the correct quark masses and mixings including
a CKM matrix that features a right-angled unitarity triangle with $\alpha\approx
\pm 90^\circ$. The positive sign of $\alpha$ is again obtained by
requiring a relative sign difference between the omitted real order one 
coefficients of the 1-2 and 2-2 elements of either $M_u$ or $M_d$ (but not
both).

Turning to the lepton sector, we point out that the
charged lepton mass matrix is related to $M_d$ by transposition and an
additional Georgi-Jarlskog factor of $-3$ in the entries that arise from the
$H_{\ol{45}}$ term of Eq.~(\ref{down1}), i.e.
\be
M_e \sim \begin{pmatrix}0 &  \mathrm{i}\lambda^{5} &0 \\
\mathrm{i}\lambda^{5}& -3 \lambda^{4}&0
\\0&-6\lambda^{4}&\lambda^{2} \end{pmatrix} v_d\ .\label{me}
\ee
Notice that, unlike in the above $A_4$ model, the left-handed 1-2 charged
lepton mixing involves a maximal phase. In the parametrisation of
\cite{sumrule1} we get for the left-handed charged lepton mixing~$V_{E_L}$
\be
\theta^{E_L}_{12} \sim \frac{\lambda}{3} \ , \qquad
\theta^{E_L}_{13} = \theta^{E_L}_{23} = 0 \ , \qquad
\phi^{E_L}_2 = \phi^{E_L}_3 = \pm \frac{\pi}{2} \ , \qquad
\chi^{E_L} = 0 \ ,
\ee
where the sign ambiguity is related to the relative sign difference between
the coefficients of the 1-2 and 2-2 elements of $M_d$ in
Eq.~(\ref{mumd}): the $+\frac{\pi}{2}$ solution corresponds to identical signs
while the $-\frac{\pi}{2}$ solution corresponds to opposite signs.
In the neutrino sector, defined by the superpotential of Eq.~(\ref{nu2})
as well as the flavon alignments of Eq.~(\ref{fnu3}), we obtain 
the Dirac and Majorana mass matrices
\be
M_D \sim \begin{pmatrix}1 &  0 &0 \\
0& 0&1\\
0&1&0 \end{pmatrix} v_u\ , \qquad
M_R \sim \begin{pmatrix} 
\alpha +2\gamma & \beta-\gamma & \beta-\gamma\\
\beta-\gamma& \beta+2\gamma&\alpha-\gamma\\
\beta-\gamma& \alpha-\gamma & \beta+2\gamma\end{pmatrix} \lambda^4 M \ ,
\ee 
where $\alpha$, $\beta$, $\gamma$ are independent order one
coefficients. The effective light neutrino mass matrix after applying the
seesaw mechanism is of exact tri-bimaximal form~\cite{Hagedorn:2010th}. This
can be easily understood as the superpotential of Eq.~(\ref{nu2}) remains
symmetric under the Klein symmetry~\cite{King:2009ap} after the flavons
$\phi^\nu$ acquire their vevs. With these vevs taking real values, the light
neutrino mass matrix ends up being purely real as well. Therefore, it is
diagonalised by the tri-bimaximal mixing matrix without any phases. The
resulting entries on the diagonal can in general be positive or negative. The
latter case would require a Majorana phase $\omega^{\nu_L}_i = \pi/2$ which,
however, does not violate CP. The neutrino mixing matrix $V_{\nu_L}$ is thus
parametrised by\footnote{The phase $\phi^{\nu_L}_3\!=\pi$ has been introduced because the
  conventional tri-bimaximal mixing matrix would otherwise  be
  incompatible with the standard PDG parametrisation where mixing angles
  take values between $0^\circ$ and~$90^\circ$.} 
\be
\sin \theta^{\nu_L}_{12} = \frac{1}{\sqrt{3}} \ , \qquad
\sin \theta^{\nu_L}_{23} = \frac{1}{\sqrt{2}} \ , \qquad
\theta^{\nu_L}_{13} = 0 \ , \qquad
\phi^{\nu_L}_2 =  \chi^{\nu_L} = 0 \ ,\qquad 
\phi^{\nu_L}_3 = \pi \ .
\ee
Using the general relations of \cite{sumrule1} it is then straightforward to
calculate the resulting MNS matrix $V_{E_L} V_{\nu_L}^\dagger$ in terms of the
two distinct left-handed mixing matrices, yielding
\bea
\sin \theta^{\text{MNS}}_{23} \, e^{-\mathrm{i}\delta_{23} } &\approx & \frac{1}{\sqrt{2}}\,
e^{-\mathrm{i}(\omega^{\nu_L}_2-\omega^{\nu_L}_3) }  \ ,\\
\theta^{\text{MNS}}_{13} \, e^{-\mathrm{i}\delta_{13} } &\approx & - \frac{\lambda}{3
  \sqrt{2}}\,e^{-\mathrm{i}(\omega^{\nu_L}_1-\omega^{\nu_L}_3 \pm\frac{\pi}{2})
}  \ , \\
\sin \theta^{\text{MNS}}_{12} \, e^{-\mathrm{i}\delta_{12} } &\approx & \frac{1}{\sqrt{3}}\,
e^{-\mathrm{i}(\omega^{\nu_L}_1-\omega^{\nu_L}_2) } 
\underbrace{\left(1\pm \mathrm{i} \frac{\lambda}{3}      \right)}_{\approx 
e^{\pm\mathrm{i}\frac{\lambda}{3} }}  \ .
\eea
This leads to a Dirac CP phase 
\be
\delta_{\mathrm{MNS}} ~=~ \delta_{13} - \delta_{23} - \delta_{12} 
~\approx~ 
\mp\left( \frac{\pi}{2} -\frac{\lambda}{3}\right)  
~\approx~\mp 86^\circ \ ,
\ee
which is maximal up to a small correction of about $4^\circ$.
As before, with tri-bimaximal neutrino mixing and charged lepton corrections,
the predictions satisfy the lepton mixing sum rule $\theta^{\text{MNS}}_{12} -
\cos(\delta_{\text{MNS}})\theta^{\text{MNS}}_{13} \approx \arcsin
(1/\sqrt{3})$ \cite{sumrule1,sumrule2}, with $\theta^{\text{MNS}}_{13}\approx \lambda /(3\sqrt{2})\approx 3^\circ$. 
In this $S_4$ model we therefore predict 
$\theta^{\text{MNS}}_{12} \approx  35.5^\circ$ 
corresponding to $\delta_{\text{MNS}} \approx \mp 86^\circ$, together with 
$\theta^{\text{MNS}}_{13}\approx 3^\circ$ and
$\theta^{\text{MNS}}_{23}\approx 45^\circ$, with an estimated error on these
predictions of $\mathcal O(1^\circ)$ or smaller attributed to renormalisation
group and canonical normalisation corrections \cite{Boudjemaa:2008jf}.

\subsection{Flavon sector}

In the following we discuss the origin of the vacuum configuration as given in
Eqs.~(\ref{fup3}-\ref{fnu3}). The flavon potential is made up of two types of
terms: $(i)$ terms which give the alignment only and $(ii)$ terms which render
the vevs real or imaginary. We first list the terms of type $(i)$ which
strongly resemble the ones used in \cite{Hagedorn:2010th}. 
\bea
W_{\mathrm{flavon}}^{(i)}&=& Y^\nu_2  \zeta^{Y^\nu_2}_{1} \frac{1}{M}  (\phi^\nu_{1}\phi^\nu_{2}+\phi^\nu_{2}\phi^\nu_{2} +\phi^\nu_{3'}\phi^\nu_{3'})
\,+\, Z^\nu_{3'}  \zeta^{Z^\nu_{3'}}_{1}\frac{1}{M}
(\phi^\nu_{1}\phi^\nu_{3'}+\phi^\nu_{2}\phi^\nu_{3'}
+\phi^\nu_{3'}\phi^\nu_{3'}) ~~~~\label{fla8} \\
&&+~X^d_1  \zeta^{X^d_1}_{1}\frac{1}{M} (\phi^d_2)^2 \phantom{\frac{1}{M}}\label{fla1}\\
&&+~ Y^d_2  \zeta^{Y^d_2}_{1}\frac{1}{M^3}(\phi^d_2)^2(\phi^d_3)^2\label{fla2}\\
&&+~\wt X^d_1  \zeta^{\wt X^d_1}_{1}\frac{1}{M^2}\phi^d_2 \phi^d_3 \wt \phi^d_3 
\,+\,\wt X^{\nu d}_{1'}  \zeta^{\wt X^{\nu d}_{1'}}_{1}\frac{1}{M^3} 
\left[(\phi^d_3)^2\right]_{3'} \phi^\nu_{3'}\wt\phi^d_3  \label{fla4}\\
&&+~ Y^{du}_2  \zeta^{Y^{du}_2}_{1} \frac{1}{M} \phi^d_2\phi^u_2\label{fla5}\\
&&+~ X^{\nu d}_{1'}  \zeta^{X^{\nu d}_{1'}}_{1}\frac{1}{M}
\phi^\nu_2 \wt\phi^d_2 \ .\phantom{\frac{1}{M}}\label{fla7}
\eea
Note that we have introduced the auxiliary $S_4$ singlet fields $\zeta^{}_{1}$
each of which being associated to a particular driving field. 
These fields allow us to impose a number of additional $\wt {\mathbb{Z}}_n$
symmetries which prove useful in the construction of a messenger completion of
the model. Under the $\wt {\mathbb{Z}}_n^{}$ symmetries, all matter, Higgs
and flavon fields are taken to be neutral. The only $\wt {\mathbb{Z}}_n^{}$
charged particles are thus the driving fields, the associated $\zeta^{}_{1}$'s
as well as the yet to be specified messenger fields. We emphasise that it is
not our aim to present the most minimal version of such a theory but rather
one possible construction that demonstrates our method.

The terms labelled~(\ref{fla8}-\ref{fla7}) give rise to all flavon alignments
but leave the overall phases undetermined. Each line determines the alignment
of a particular flavon field successively: the terms~(\ref{fla8}) give the
$\phi^\nu$ alignments, the term labelled~(\ref{fla1}) determines $\langle
\phi^d_2 \rangle$, with this~(\ref{fla2}) fixes $\langle \phi^d_3 \rangle$, etc. As most of
the operators are taken from \cite{Hagedorn:2010th} we do not spell out the 
corresponding $F$-term conditions that lead to flavon alignments which are
identical in both scenarios. Instead we focus on the two operators of
$W_{\mathrm{flavon}}^{(i)}$ that are new in our setup: ($a$) the second term
of~(\ref{fla4}) as well as ($b$) the term labelled~(\ref{fla7}).

\begin{itemize}
\item[($a$)] As has been discussed in \cite{Hagedorn:2010th}, the first term
 of~(\ref{fla4}) renders $\langle \wt \phi^d_{3,1} \rangle =
 0$. Inserting this condition as well as the alignments of $\phi^d_{3}$ and
 $\phi^\nu_{3'}$ into the second term of~(\ref{fla4}) results in
\be
\wt X^{\nu d}_{1'}  \zeta^{\wt X^{\nu d}_{1'}}_{1}\frac{1}{M^3} 
\langle \phi^d_{3,2} \rangle ^2
\langle \phi^\nu_{3',i=1,2,3} \rangle 
\left( 2 \langle \wt \phi^d_{3,3}\rangle -\langle \wt
 \phi^d_{3,2}\rangle\right) \ .
\ee
Under the assumption that the $\zeta^{}_{1}$ fields get a non-vanishing vev,
the $F$-term equation for $\wt X^{\nu d}_{1'}$ aligns $\wt \phi^d_3$ such that 
\be
\langle \wt \phi^d_3 \rangle  \propto \begin{pmatrix} 0 \\ 2
 \\1 \end{pmatrix}\ .
\ee
\item[($b$)] Plugging the $\phi^\nu_2$ alignment into the term
  labelled~(\ref{fla7}) leads to  
\be
X^{\nu d}_{1'}  \zeta^{X^{\nu d}_{1'}}_{1}\frac{1}{M}
\langle \phi^\nu_{2,i=1,2} \rangle 
\left(
\langle \wt\phi^d_{2,2} \rangle -  \langle \wt\phi^d_{2,1}
\right) \ ,
\ee
which in turn generates the desired $\wt\phi^d_2$ alignment
\be
\langle \wt \phi^d_2 \rangle  \propto \begin{pmatrix} 1 \\ 1 \end{pmatrix}\ .
\ee
\end{itemize}

Let us now turn to the second type of flavon potential terms. In order to fix
the  phases of all flavon  vevs we introduce the extra flavons $\xi_1$ and
$\wt\xi_{1'}$. The part of the flavon potential that renders the pre-aligned
vevs either real or imaginary then reads 
\bea
W_{\mathrm{flavon}}^{(ii)}&=&
P_0^{(1)}  \,\zeta^{P_0^{(1)}}_{1} \left[\frac{1}{M}(\xi_1)^2 - m^{(1)}
\right]
~+~P_0^{(2)}  \,\zeta^{P_0^{(2)}}_{1} \left[\frac{1}{M}(\wt\xi_{1'})^2- m^{(2)}\right] \label{real1new}\\
&&
+~P_0^{(3)}  \,\zeta^{P_0^{(3)}}_{1} \left[\frac{1}{M}(\phi^\nu_1)^2- m^{(3)}\right] \\
&&+~ P_1^{(1)} \,\zeta^{P_1^{(1)}}_{1}\left[\frac{1}{M}(\phi^u_{1'})^2 - c^{(1)} \, \xi_1\right]
~+~ P_1^{(2)} \,\zeta^{P_1^{(2)}}_{1}\left[\frac{1}{M}(\wt\phi^d_{2})^2  - c^{(2)} \, \xi_1\right]\\
&&+~ P_1^{(3)} \,\zeta^{P_1^{(3)}}_{1}\left[\frac{1}{M}(\wt\phi^d_{3})^2  - c^{(3)} \, \xi_1\right]
~+~ P_1^{(4)} \,\zeta^{P_1^{(4)}}_{1}\left[ \frac{1}{M^2}(\phi^d_{2})^2 \phi^\nu_2 - c^{(4)} \, \xi_1\right]~~~~~~~~~\label{real2new}\\ 
&&+~\wt P_{1'}^{(1)} \,\zeta^{\wt P_{1'}^{(1)}}_{1}\left[\frac{1}{M^2}(\phi^d_{3})^2  \phi^\nu_2  - \wt c^{(1)} \, \wt\xi_{1'} \right] \\
&&+~\wt P_{1'}^{(2)} \,\zeta^{\wt P_{1'}^{(2)}}_{1} \wt \zeta^{\wt
 P_{1'}^{(2)}}_{1}\frac{1}{M} \left[
\frac{1}{M^4}\phi^u_2 (\phi^d_{2})^4 - \wt c^{(2)} \, \wt\xi_{1'} \right]
\label{real3new}  \ .~~~~~~~~~~~ 
\eea 
As before, each of the driving fields $P_0^{(i)}$, $P_1^{(i)}$, $\wt
P_{1'}^{(i)}$ has an associated $\zeta^{}_{1}$ field which allows us to
segregate the messengers of the different effective terms.\footnote{In the
  case of $\wt P_{1'}^{(2)}$, it is possible to show that this separation only
  works if one introduces two $\zeta^{}_{1}$ fields.}  
In order to get the $\lambda$ suppressions of the flavons as given in
Eqs.~(\ref{fup3}-\ref{fnu3}) we can choose the parameters
\be
m^{(i)} ~\sim~  \lambda^8 M 
\qquad \rightarrow  \qquad
\langle \xi_1 \rangle  ~\sim~ \langle \wt \xi_{1'} \rangle ~\sim~ \lambda^4 M
\ ,
\ee
\be
c^{(i)} ~\sim~ \lambda^2  \ , \qquad
\wt c^{(i)} ~\sim~ \lambda^4  \ .
\ee
We emphasise that this hierarchy in the parameters $\frac{m^{(i)}}{M}$,
$c^{(i)}$ and $\wt c^{(i)}$ is an ad hoc assumption but a necessity in the
setup of a model with hierarchical flavon vevs.

With the $\zeta_1$ fields acquiring non-vanishing (and possibly complex)
vevs, the $F$-term equations of the driving fields in
$W^{(ii)}_{\mathrm{flavon}}$ determine the phases of the pre-aligned flavon
vevs. Since we require CP conservation of our underlying theory, all coupling
and mass parameters can be taken real. Assuming that all
parameters -- $m^{(i)}$, $\langle \xi_1\rangle $, $\langle\wt \xi_{1'}\rangle$,
$c^{(i)}$, $\wt c^{(i)}$, as well as the suppressed order one coefficients -- are
positive, it is straightforward to see that all flavon vevs turn out to be
real. If we now flip the sign of the parameter $c^{(2)}$ while keeping
everything else unchanged, the vev of $\wt \phi^d_2$ is driven to a purely
imaginary value, as required by Eq.~(\ref{fdown4}). As a final remark, we
mention that only one of the neutrino-type flavon vevs has to be driven to a
real value. This is a consequence of the alignment terms~(\ref{fla8})
which relate the three vevs in a simple way, cf. \cite{Hagedorn:2010th}.

\subsection{All possible $\bs{\mathbb{Z}_n}$ symmetries}

Having fixed our desired superpotential operators, we now determine the maximal
set of $\mathbb Z_n$ symmetries that is consistent with the effective terms~(\ref{up1}-\ref{nu2},\ref{fla8}-\ref{fla7},\ref{real1new}-\ref{real3new}). 
We do this in order to check whether or not we need to worry about additional
operators that might spoil our model at the effective
level. In~\cite{Hagedorn:2010th}, a $U(1)$ symmetry was invoked to forbid such
dangerous terms; due to the structure of the superpotential
terms~(\ref{real1new}-\ref{real3new}) a $U(1)$ symmetry is not possible here
and we have to confine ourselves to $\mathbb Z_n$ symmetries. Defining
the parameters $a_i=0,1$ we immediately obtain several $\mathbb Z_n$ charges
$z[\text{field}]$ from the terms labelled~(\ref{real1new}-\ref{real3new}),
\be
z[\xi_1] \,=\, a_1\frac{n}{2}    \ ,\qquad
z[\wt \xi_{1'}] \,=\, a_{1'} \frac{n}{2}    \ ,\qquad
z[\phi^\nu] \,=\, a_2 \frac{n}{2}   \ ,
\ee
\be
z[\phi^u_{1'}] \,=\,  \left(\frac{a_1}{2} + a_3 \right) \frac{n}{2}\ ,
\qquad
z[\wt \phi^d_2] \,=\, \left(\frac{a_1}{2} + a_4 \right) \frac{n}{2} \ ,\qquad
z[\wt \phi^d_3] \,=\, \left(\frac{a_1}{2} + a_5 \right) \frac{n}{2} \
,\label{znfirst} 
\ee
\be
z[\phi^d_2] \,=\, \left(\frac{a_1-a_2}{2} + a_6 \right) \frac{n}{2} \ ,
\ee
\be
z[\phi^d_3] \,=\, \left(\frac{a_{1'}-a_2}{2} + a_7 \right) \frac{n}{2}\ ,\qquad
z[\phi^u_2] \,=\,  a_{1'} \frac{n}{2}\ .
\ee 
Turning to the Yukawas of Eqs.~(\ref{up1}-\ref{nu2}) yields the $\mathbb Z_n$
charges of the remaining fields. In the up sector, the existence of the first
two non-renormalisable terms demands $z[\phi^u_2] = 2z[\phi^u_{1'}]$, leading to
$a_1=a_{1'}$. Introducing the integer parameter $\alpha=0,...,n-1$, we find
the relations
\be
z[T_3] \,=\, \alpha \ , \qquad
z[H_{5}] \,=\, -2\alpha \ ,  \qquad
z[T] \,=\, \alpha +\left( -\frac{a_1}{2} - a_{4'}\right) \frac{n}{2} \ ,
\ee
\be
z[N] \,=\,  \left(-\frac{a_2}{2} +a_8 \right) \frac{n}{2}\ ,\qquad
z[F] \,=\,  2\alpha + \left(\frac{a_2}{2} - a_8 \right) \frac{n}{2} \ , 
\ee
\be
z[H_{\ol{5}}] \,=\, -3\alpha + \left(- \frac{a_1}{2} -a_7+a_8 \right) \frac{n}{2} \ ,
\ee
\be
z[H_{\ol{45}}] \,=\, -3\alpha + \left( -\frac{a_1}{2} +a_{4'}-a_5-a_6+a_8 \right)\frac{n}{2}  \ .\label{znlast}
\ee
Here the charge of $H_{\ol{5}}$ is calculated from the first term of
Eq.~(\ref{down1}). In order for the third term of Eq.~(\ref{down1}) to be
allowed as well, we need to set $a_{4'}=a_4$.

This leaves us with eight parameters $a_i=0,1$ and one integer parameter
$\alpha$. 
Choosing {\it all but one of these parameters non-zero} defines a particular
$\mathbb Z_n^{(i)}$ symmetry. The resulting set of $\mathbb Z_n^{(i)}$ symmetries can then be
used to constrain the allowed terms of our model. Let us identify these
initial $\mathbb Z_n^{(i)}$ symmetries.

Setting all $a_i=0$ and keeping only the integer parameter $\alpha$, we obtain
a symmetry with neutral flavon fields and non-zero charges for the matter and
Higgs fields,
$$
z[T_3]=z[T]=\alpha \ , ~~~
z[H_5]= -2\alpha \ ,~~~
z[N]= 0 \ ,~~~
z[F]=2\alpha \ , ~~~
z[H_{\ol{5}}]=z[H_{\ol{45}}]= -3\alpha \ .
$$
Such a symmetry is, however, always respected for $SU(5)$ invariant products of
two matter and one Higgs field and therefore not useful in constraining our
model. Hence we can simply disregard such a symmetry.

It is clear that the remaining eight parameters can only give rise to either
$\mathbb Z_2$ or $\mathbb Z_4$ symmetries. The $\mathbb Z_2$ symmetry derived from setting all
parameters to zero except for $a_8=1$ yields non-vanishing charge only for $N$,
$F$, $H_{\ol{5}}$, and $H_{\ol{45}}$. Such a symmetry is, again, respected for
all $SU(5)$ invariant terms so that it is not helpful in constraining our
model. Hence we also disregard this symmetry.

This leaves us with seven useful symmetries, five $\mathbb Z_2$'s and two
$\mathbb Z_4$'s. They are summarised in Tab.~\ref{ZnsMatterHiggsFlavon}. 
\begin{table}[t]
\centering
$$
\begin{array}{ccccccccccc}\toprule
& SU(5) & S_4 & U(1)_R & {\mathbb{Z}}^{(1)}_4 & {\mathbb{Z}}^{(2)}_4 &
{\mathbb{Z}}^{(3)}_2 & {\mathbb{Z}}^{(4)}_2 & {\mathbb{Z}}^{(5)}_2 &
{\mathbb{Z}}^{(6)}_2 & {\mathbb{Z}}^{(7)}_2 \\\midrule 
\multicolumn{11}{l}{\text{Matter~Fields}} \\ \midrule
T            &{\bf 10}     &{\bf 2}      & 1 & 3 & 0 & 0 & 1 & 0 & 0 & 0    \\ 
T_3          &{\bf 10}     &{\bf 1}      & 1 & 0 & 0 & 0 & 0 & 0 & 0 & 0    \\ 
F            &{\bf \ol 5}  &{\bf 3}      & 1 & 0 & 1 & 0 & 0 & 0 & 0 & 0    \\ 
N            &{\bf 1}      &{\bf 3}      & 1 & 0 & 3 & 0 & 0 & 0 & 0 & 0    \\
\midrule
\multicolumn{11}{l}{\text{Higgs~Fields}} \\ \midrule
H_5          &{\bf 5}      &{\bf 1}      & 0 & 0 & 0 & 0 & 0 & 0 & 0 & 0    \\ 
H_{\ol{5}}   &{\bf \ol 5}  &{\bf 1}      & 0 & 3 & 0 & 0 & 0 & 0 & 0 & 1    \\ 
H_{\ol{45}}  &{\bf \ol{45}}&{\bf 1}      & 0 & 3 & 0 & 0 & 1 & 1 & 1 & 0    \\ \midrule
\multicolumn{11}{l}{\text{Flavons}} \\ \midrule
\xi_1        &{\bf 1}     &{\bf 1}      & 0 & 2 & 0 & 0 & 0 & 0 & 0 & 0    \\ 
\wt\xi_{1'}  &{\bf 1}     &{\bf 1'}     & 0 & 2 & 0 & 0 & 0 & 0 & 0 & 0    \\[1mm]
\phi^u_2     &{\bf 1}     &{\bf 2}      & 0 & 2 & 0 & 0 & 0 & 0 & 0 & 0    \\ 
\phi^u_{1'}  &{\bf 1}     &{\bf 1'}     & 0 & 1 & 0 & 1 & 0 & 0 & 0 & 0    \\[1mm]
\phi^d_3     &{\bf 1}     &{\bf 3}      & 0 & 1 & 3 & 0 & 0 & 0 & 0 & 1    \\ 
\wt\phi^d_3  &{\bf 1}     &{\bf 3}      & 0 & 1 & 0 & 0 & 0 & 1 & 0 & 0    \\ 
\phi^d_2     &{\bf 1}     &{\bf 2}      & 0 & 1 & 3 & 0 & 0 & 0 & 1 & 0    \\
\wt\phi^d_2  &{\bf 1}     &{\bf 2}      & 0 & 1 & 0 & 0 & 1 & 0 & 0 & 0    \\[1mm]
\phi^\nu     &{\bf 1}     &{\bf 3',2,1} & 0 & 0 & 2 & 0 & 0 & 0 & 0 & 0    \\ \bottomrule
\end{array}
$$
\caption{\label{ZnsMatterHiggsFlavon}All possible symmetries of the effective
 $S_4$ model with matter, Higgs and flavon fields. These fields are assumed
 to be neutral under the additional 17 $\wt{\mathbb{Z}}_n$ symmetries.}
\end{table}
Note that this set of symmetries forbids the bilinear term $NN$.
We also remark that all flavon fields are distinguished from one another by
their $\mathbb Z_n$ charges. Similarly, assuming the $\zeta^{}_{1}$ fields to
be neutral under the above seven $\mathbb Z_n$ symmetries, all driving fields
in~(\ref{fla8}-\ref{fla7}) are charged differently except for $Y^\nu_2$,
$Z^\nu_{3'}$ and $Y^d_2$ which are neutral. In order to obtain the desired
flavon alignment we need to distinguish $Y^d_2$ from $Y^\nu_2$ with some
quantum number. This is achieved by introducing the auxiliary fields
$\zeta^{Y^d_2}_{1}$ and $\zeta^{Y^\nu_2}_{1}$. In addition we impose two new
$\wt {\mathbb Z}_n$ symmetries such that $\zeta^{Y^d_2}_{1}$ is only charged
under the first $\wt {\mathbb Z}_n$ while $\zeta^{Y^\nu_2}_{1}$ sees only the
second $\wt {\mathbb Z}_n$. The matter, Higgs, and flavon fields are all
assumed to be neutral under these $\wt {\mathbb Z}_n$ symmetries. However, the
driving fields do carry a $\wt {\mathbb Z}_n$ charge such that it compensates
the charge of the corresponding $\zeta_1$ field. Thus it is possible to
distinguish $Y^d_2$ from $Y^\nu_2$. 
Even though $Y^d_2\zeta^{Y^d_2}_{1}$ and $Y^\nu_2\zeta^{Y^\nu_2}_{1}$ have 
identical net quantum numbers, it is possible to argue that the former
couples to $(\phi^d_2)^2(\phi^d_3)^2$, as shown in (\ref{fla2}),
while the latter couples to $(\phi^\nu_{})^2$, see the terms of~(\ref{fla8}). 
We will get back to this example in Sec.~\ref{sec:s4allowedterms}. 

The procedure of introducing a new $\wt {\mathbb Z}_n$ symmetry for each
driving field and its associated $\zeta_{1}$ field leads to a total of 17
symmetries. In the high energy completion of our model we choose 16 $\wt
{\mathbb Z}^{(k)}_2$ and one $\wt {\mathbb Z}_4$ symmetry. As already
mentioned the matter, Higgs and flavon fields are neutral under these
additional symmetries, so that a driving field and its associated $\zeta_{1}$
have opposite $\wt {\mathbb Z}_n$ charges. Tab.~\ref{Znsdriving} lists the
charges of the driving fields under all possible symmetries. 
\begin{table}[t]
\centering
$$
\begin{array}{ccccccccccccc}\toprule
\text{Driving Fields} & SU(5) & S_4 & U(1)_R & {\mathbb{Z}}_4^{(1)} & {\mathbb{Z}}_4^{(2)} & {\mathbb{Z}}_2^{(3)} &
{\mathbb{Z}}_2^{(4)} & {\mathbb{Z}}_2^{(5)} & {\mathbb{Z}}_2^{(6)} & {\mathbb{Z}}_2^{(7)}  
& \wt {\mathbb{Z}}_2^{(k)} & \wt {\mathbb{Z}}_4^{(17)} \\\midrule
{X^d_1} & {\bf 1} & {\bf 1} & 2 & 2 & 2 & 0 & 0 & 0 & 0 & 0 & \delta_{1k} & 0 \\
{Y^d_2} & {\bf 1} & {\bf 2} & 2 & 0 & 0 & 0 & 0 & 0 & 0 & 0 & \delta_{2k} & 0 \\
\wt X^d_1 & {\bf 1} & {\bf 1} & 2 & 1 & 2 & 0 & 0 & 1 & 1 & 1 & \delta_{3k} & 0 \\
\wt X^{\nu d}_{1'} & {\bf 1} & {\bf 1'} & 2 & 1 & 0 & 0 & 0 & 1 & 0 & 0 & \delta_{4k} & 0 \\
Y^{du}_2 & {\bf 1} & {\bf 2} & 2 & 1 & 1 & 0 & 0 & 0 & 1 & 0 & \delta_{5k} & 0 \\
X^{\nu d}_{1'} & {\bf 1} & {\bf 1'} & 2 & 3 & 2 & 0 & 1 & 0 & 0 & 0 & \delta_{6k} & 0 \\
Y^\nu_2 & {\bf 1} & {\bf 2} & 2 & 0 & 0 & 0 & 0 & 0 & 0 & 0 & \delta_{7k} & 0 \\
Z^\nu_{3'} & {\bf 1} & {\bf 3'} & 2 & 0 & 0 & 0 & 0 & 0 & 0 & 0 & \delta_{8k} & 0\\\midrule
P_0^{(1)} & {\bf 1} & {\bf 1} & 2 & 0 & 0 & 0 & 0 & 0 & 0 & 0 & \delta_{9k}  & 0\\
P_0^{(2)} & {\bf 1} & {\bf 1} & 2 & 0 & 0 & 0 & 0 & 0 & 0 & 0 & \delta_{10\,k} & 0 \\
P_0^{(3)} & {\bf 1} & {\bf 1} & 2 & 0 & 0 & 0 & 0 & 0 & 0 & 0 & \delta_{11\,k}  & 0\\
P_1^{(1)} & {\bf 1} & {\bf 1} & 2 & 2 & 0 & 0 & 0 & 0 & 0 & 0 &
\delta_{12\,k} & 0 \\
P_1^{(2)} & {\bf 1} & {\bf 1} & 2 & 2 & 0 & 0 & 0 & 0 & 0 & 0 & \delta_{13\,k}  & 0\\
P_1^{(3)} & {\bf 1} & {\bf 1} & 2 & 2 & 0 & 0 & 0 & 0 & 0 & 0 & \delta_{14\,k} & 0 \\
P_1^{(4)} & {\bf 1} & {\bf 1} & 2 & 2 & 0 & 0 & 0 & 0 & 0 & 0 & \delta_{15\,k}  & 0\\
\wt P_{1'}^{(1)} & {\bf 1} & {\bf 1'} & 2 & 2 & 0 & 0 & 0 & 0 & 0 & 0 & \delta_{16\,k}  & 0\\
\wt P_{1'}^{(2)} & {\bf 1} & {\bf 1'} & 2 & 2 & 0 & 0 & 0 & 0 & 0 & 0 & 0 & 1 \\\bottomrule
\end{array}
$$
\caption{\label{Znsdriving}The charges of the driving fields under all
 possible symmetries in the $S_4$ model. The 16 $\wt{\mathbb Z}_2$ symmetries
 are enumerated by $k=1,...,16$ and $\delta_{ik}$ denotes the Kronecker
 delta. Hence, $X^d_1$, e.g.\ has a $\wt{\mathbb Z}_2^{(1)}$ charge of 1 and is
 neutral under the remaining 15 $\wt{\mathbb Z}_2^{(k)}$ symmetries.}
\end{table}

\subsection{\label{sec:s4allowedterms}Effectively allowed terms and
 messengers} 

Having imposed the maximal set of symmetries we can forbid many unwanted
terms. However, it needs to be checked whether or not these symmetries are
powerful enough to forbid {\it all} unwanted operators. We therefore
determine the additionally allowed terms at the effective level with equal or
less $\lambda$ suppression compared to the desired ones. 

In the Yukawa sector we obtain no additional such terms for $W_d$ and
$W_\nu$. The only unwanted terms arise in the up sector, namely
\bea
\Delta W_u&=& TT_3H_5 \,{\frac{1}{M} \wt \phi^d_2} 
~+~ TTH_5 \left[ {\frac{1}{M}\xi_1 + \frac{1}{M^2}(\wt\phi^d_3)^2 +
 \frac{1}{M^2}(\wt\phi^d_2)^2 + \frac{1}{M^3}(\phi^d_2)^2\phi^\nu_{}} \right]
.~~~\label{danger-up}
\eea
The first term of Eq.~(\ref{danger-up}) leads to entries of order $\lambda^3$
in the $T_1 T_3$ as well as the $T_2T_3$ element of the up quark Yukawa
matrix. At the effective level this term is unavoidable if we require the
first and third term of the Yukawa couplings of Eq.~(\ref{down1}) together
with the first term of Eq.~(\ref{up1}). 
The first term in the bracket of Eq.~(\ref{danger-up}) is of order $\lambda^4$
and contributes to the 1-2 element, while the remaining three terms are
$\lambda^6$ suppressed and contribute to the 1-1 element. 

Turning to the flavon potential, the additionally allowed effective terms read
\bea
\Delta W_{\mathrm{flavon}}^{(i)}&=&
Y^\nu_2 \zeta^{Y^\nu_2}_{1} \frac{1}{M}  \left[ {\frac{1}{M^2}\left[ (\phi^d_3)^4 +(\phi^d_2)^2(\phi^d_3)^2+(\phi^d_2)^4\right]+ (\phi^u_2)^2 +  \phi^u_2(\xi_1  + \wt \xi_{1'})} \right]\label{mess-begin}\\
&& +~ Z^\nu_{3'}  \zeta^{Z^\nu_{3'}}_{1} \frac{1}{M^3} \left[{
   (\phi^d_3)^4 + (\phi^d_2)^2(\phi^d_3)^2}   \right]\label{mess1a}\\
&& +~Y^d_2  \zeta^{Y^d_2}_{1}\frac{1}{M^3}  {(\phi^d_2)^4} \label{mess1b}\\
&& +~\wt X^{\nu d}_{1'}\zeta^{\wt X^{\nu d}_{1'}}_{1} \frac{1}{M^2}
\left[{(\wt\phi^d_3)^3 + \frac{1}{M}\wt\phi^d_3 (\phi^d_2)^2\phi^\nu_{} }
\right] \\ 
&& +~ Y^{du}_2  \zeta^{Y^{du}_2}_{1} \frac{1}{M} {  \phi^d_2 (\xi_1+ \wt \xi_{1'})} \ ,\label{mess2}\\[2mm]
\Delta W_{\mathrm{flavon}}^{(ii)}&=&\sum_{i=1}^{3} P_0^{(i)}
\zeta^{P_0^{(i)}}_{1} \frac{1}{M^3} {(\phi^d_3)^2(\phi^d_2)^2} \\
&& +~ 
\left( \wt P_{1'}^{(1)}  \zeta^{\wt P_{1'}^{(1)}}_{1}  \frac{1}{M^2} +
\wt P_{1'}^{(2)}  \zeta^{\wt P_{1'}^{(2)}}_{1} \wt \zeta^{\wt P_{1'}^{(2)}}_{1} \frac{1}{M^3}
\right) {(\phi^d_{2})^2  \phi^\nu_{}}  \ .\label{mess-end}
\eea

We see that it is impossible to formulate the model consistently at the
effective level. However, as all of the above unwanted terms are
non-renormalisable, we can forbid them in a high energy completion of the
model with suitably chosen messengers. We have worked out explicitly that such
a completion can be realised straightforwardly, the details of which are
presented in App.~\ref{app-messenger}. Here we simply want to illustrate
our strategy which employs the set of $\wt {\mathbb Z}_n$ symmetries and the
associated $\zeta^{}_{1}$ fields on two examples.

\begin{figure}
\begin{center}
\includegraphics[width=14cm]{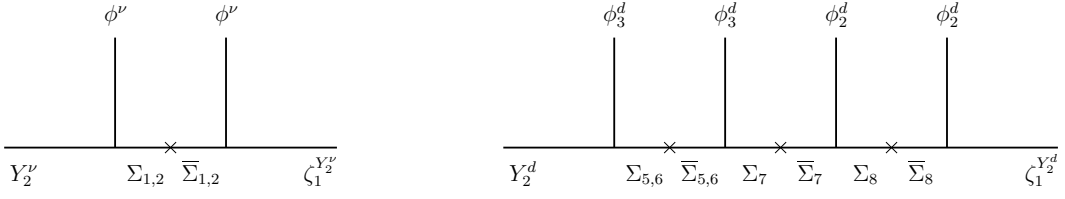}
\end{center}
\caption{\label{fig:mess4}The underlying diagrams for the effective flavon
 superpotential terms~(\ref{fla8},\ref{fla2}) with the driving
 fields $Y^\nu_2$ and $Y^d_2$.}
\end{figure}

As mentioned previously, the driving fields $Y^\nu_2$ and $Y^d_2$ can be
distinguished by introducing the two auxiliary fields $\zeta^{Y^\nu_2}_{1}$
and $\zeta^{Y^d_2}_{1}$. Then the underlying diagrams that give rise to the
corresponding effective flavon superpotential terms~(\ref{fla8},\ref{fla2})
are given as shown in Fig.~\ref{fig:mess4}.  
The set of $\wt {\mathbb Z}_n$ symmetries separates the messengers of both
diagrams. While $\Sigma_{1,2}$, $\ol\Sigma_{1,2}$ are only charged under $\wt
{\mathbb Z}_2^{(7)}$, the messengers in the right diagram $\Sigma_{5,6,7,8}$,
$\ol\Sigma_{5,6,7,8}$ carry non-trivial charge only under  $\wt {\mathbb
 Z}_2^{(2)}$. Thus the messenger in the left diagram cannot appear
in the right diagram and vice versa so that at the effective level, $Y^\nu_2
\zeta^{Y^\nu_2}_{1}$ cannot couple to $(\phi^d_2)^2(\phi^d_3)^2$ although
$Y^d_2\zeta^{Y^d_2}_{1}$ does.

\begin{figure}
\begin{center}
\includegraphics[width=14cm]{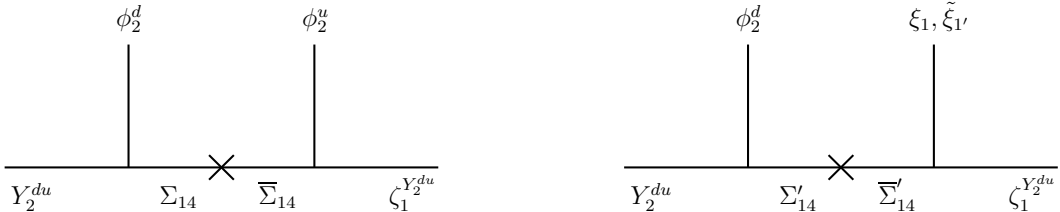}
\end{center}
\caption{\label{fig:mess3}On the left, the underlying diagram for the
 effective flavon superpotential term~(\ref{fla5}) with the driving
 field~$Y^{du}_2$. On the right: a possible diagram for the unwanted terms
 of~(\ref{mess2}).} 
\end{figure}

As our second example let us consider the effective terms involving the
driving field $Y^{du}_2$, i.e. the terms of~(\ref{fla5}) and (\ref{mess2}). The
underlying structure for the former operator is presented in the left diagram
of Fig.~\ref{fig:mess3}.
The corresponding diagrams for the unwanted effective terms are shown on the
right. Clearly, in both cases the messengers have identical $\wt {\mathbb Z}_n$
charges. However, while $\Sigma_{14}$ is an $S_4$ doublet, the messenger
$\Sigma'_{14}$ would have to furnish a one-dimensional $S_4$ representation in
order to allow the diagrams on the right. Demanding the  existence of the
doublet messenger and the absence of a similar one-dimensional one, we do not
generate the unwanted operators of~(\ref{mess2}).

Likewise, we have checked that the messenger sector of
App.~\ref{app-messenger}, which gives rise to the desired effective terms, 
does not generate any of the unwanted terms of
$\Delta W_u$, $\Delta W_{\mathrm{flavon}}^{(i)}$, $\Delta
W_{\mathrm{flavon}}^{(ii)}$. This can be easily verified by studying the
allowed renormalisable superpotential terms of the ultraviolet completed model
as given in App.~\ref{app-messenger}.

\section{Summary and conclusions}
In this paper we have proposed new classes of models which predict both
tri-bimaximal lepton mixing and a right-angled CKM unitarity triangle, $\alpha
\approx 90^\circ$. The ingredients of the models include a SUSY GUT such as
$SU(5)$ and a discrete family symmetry such as $A_4$ or $S_4$, which are
familiar ingredients of models which give rise to tri-bimaximal mixing.

The main additional restriction we impose is on the form of the shaping
symmetry which we require to consist of products of $\mathbb{Z}_2$ and
$\mathbb{Z}_4$ groups, and also the assumption of spontaneous CP violation. We
have shown how the vacuum alignment in such models allows a simple explanation
of $\alpha \approx 90^\circ$ by a combination of purely real or purely
imaginary vevs of the flavons responsible for family symmetry breaking. 
We emphasise that 
the approach we have proposed is based on a general method for the vacuum alignment of the flavon fields with additional discrete $\mathbb{Z}_n$ shaping symmetries, which forces the phases of the flavon vevs to take only discrete values. For the special case of  $\mathbb{Z}_2$ and $\mathbb{Z}_4$ symmetries, the vevs of the flavon fields can be forced to be purely real or purely imaginary.

Another requirement is that the models must lead to quark mass matrices
with 1-3 texture zeros in order to satisfy the ``phase sum rule''
$\alpha \approx \delta_{12}^d - \delta_{12}^u$ \cite{Antusch:2009hq}, where the phases $\delta_{12}^d$ and $\delta_{12}^u$ are the arguments of the complex 1-2 rotation angles in the up-type and down-type quark mass matrices. To explain $\alpha \approx  90^\circ$ one might therefore simply try to realise $\delta_{12}^d =  90^\circ$, $\delta_{12}^u = 0$ or alternatively $\delta_{12}^d = 0$, $\delta_{12}^u = - 90^\circ$ in a model of flavour.
The lepton mass matrices also satisfy the ``lepton mixing sum rule'' together with a new prediction that the 
leptonic CP violating oscillation phase is close to either $0^\circ$, $90^\circ$,
$180^\circ$, or $270^\circ$ depending on the model,
with neutrino masses being purely real (no complex Majorana phases).
This leads to the possibility of having right-angled unitarity triangles in both the quark and lepton sectors.

We have constructed two explicit $SU(5)$ SUSY GUT models with $A_4$ and $S_4$ family symmetries, respectively, plus $\mathbb{Z}_n$ (even $n$) shaping symmetries in order to apply and illustrate our idea. 
The $A_4\times SU(5)$ and $S_4\times SU(5)$ models provide examples of an indirect and direct model,
with each model being a variation on a previous
model proposed in the literature, but including the above restriction on the shaping symmetry,
and also that of having 1-3 texture zeros.

In addition to the main theme of the paper, namely to realise a right-angled CKM unitarity triangle with $\alpha \approx 90^\circ$, we have found the following interesting by-product: 
in models with $S_4$ family symmetry a flavon with a vev proportional to
$(0,2,1)$ can emerge from the vacuum alignment and could significantly improve
the prediction of the model with respect to the quark mixing angle
$\theta_{23}^{\text{CKM}}$. In our example $S_4$ model, this specific flavon vev led to the prediction $\theta_{23}^{\text{CKM}} = 2 m_s/m_b$, which is in good agreement with current experimental data. 

In summary, we have proposed a simple way to construct models that not only fit the amount of quark CP violation but which instead feature a right-angled CKM unitarity triangle with $\alpha \approx 90^\circ$, as suggested by the recent experimental data, as a prediction. The two explicit models we constructed with $A_4$ and $S_4$ family symmetries and $SU(5)$ SUSY GUTs, make predictions for the leptonic Dirac CP phase of $\delta_{\text{MNS}} \approx 0,180^\circ$ and $\delta_{\text{MNS}} \approx \pm 90^\circ$ (respectively). Furthermore, both models predict $\theta^{\text{MNS}}_{23}\approx 45^\circ$ and $\theta^{\text{MNS}}_{13}\approx 3^\circ$. The sum rule $\theta^{\text{MNS}}_{12} - \cos(\delta_{\text{MNS}})\theta^{\text{MNS}}_{13} \approx \arcsin (1/\sqrt{3})$ relates the reactor and the solar mixing angles via the Dirac CP phase.  As a result, we obtain a shift of $\theta^{\text{MNS}}_{12}$ from its tri-bimaximal value of $35.26^\circ$ which is of the order of $3^\circ$ in the model with $A_4$ family symmetry, corresponding to $\delta_{\text{MNS}} = 0^\circ,180^\circ$. In contrast, the $S_4$ model predicts $\theta^{\text{MNS}}_{12} \approx  35.5^\circ$  corresponding to $\delta_{\text{MNS}} \approx \mp 86^\circ$. These predictions are testable in future neutrino oscillation facilities \cite{Antusch:2007rk}, and the required ingredients to completely reconstruct the leptonic unitarity triangles have been discussed in~\cite{AguilarSaavedra:2000vr}. The $S_4\times SU(5)$ SUSY GUT of Flavour illustrates the interesting possibility of having right-angled unitarity triangles in both the quark and lepton sector.

\section*{Acknowledgments}
We thank Claudia Hagedorn and Graham G. Ross for helpful conversations. 
S.~A.\ acknowledges partial support by the DFG cluster of excellence ``Origin and Structure of the Universe.'' S.~F.~K.\ and C.~L.\ acknowledge support from the STFC Rolling Grant No. ST/G000557/1. S.~F.~K.\ is grateful to the Royal Society for a Leverhulme Trust Senior Research Fellowship. S.~A., S.~F.~K.\ and C.~L.\ thank the CERN theory group and the organisers of the CERN Theory Institute $\nu$TheME, where part of this work was carried out. C.~L.\ thanks the Max-Planck-Institut f\"ur Physik for hospitality. M.~S.\ acknowledges partial support from the Italian government under the project number PRIN 2008XM9HLM ``Fundamental interactions in view of the Large Hadron Collider and of astro-particle physics''.

\section*{Appendix}

\begin{appendix}

\section{\label{app-basis}The Basis of neutral driving fields}

In this appendix we want to discuss how to disentangle the various couplings of the neutral driving fields determining the phases of the flavons in the $A_4$ model by going to a suitable basis. For simplicity we want to assume for the moment that the flavons have only a $\mathbb{Z}_2$ charge and we discuss only the superpotential terms which fix the phase of the flavon vevs. With only one flavon we then have a superpotential of the kind
\begin{equation}
W = P_A (g_1 \phi_1^2 + M_A^2 ) \;,
\end{equation}
where $g_1$ is a real coupling constant which can have a-priori either sign. $M_A$ is the mass scale of the flavon field $\phi_1$. Here we are already in the suitable basis which consists of $P_A$ and the corresponding $F$-term condition fixes the phase of $\langle \phi_1 \rangle$.

If we have a second flavon $\phi_2$ with a second driving field $P_B$ the above superpotential is extended to
\begin{equation}
W = P_A (g_1 \phi_1^2 + g_2 \phi_2^2 + M_A^2 ) + P_B (h_1 \phi_1^2 + h_2 \phi_2^2 + M_B^2) \;,
\end{equation}
where $g_i$ and $h_i$ are real coupling constants and $M_A$ and $M_B$ are real mass parameters. Since the $P_A$ and $P_B$ are neutral under the discrete symmetries they can, in principle, couple to both flavon fields. In this basis the $F$-term conditions do not give obviously the desired result.
But if we assume that the coupling matrix is non-singular, we can apply the real redefinitions
\begin{equation}
P_A =  \frac{h_2 P_1 - h_1 P_2}{g_1 h_2 - g_2 h_1} \quad \text{and} \quad P_B =  \frac{g_1 P_2 - g_2 P_1}{g_1 h_2 - g_2 h_1} \;,
\end{equation}
to the superpotential, which is expressed in terms of the new fields $P_1$ and $P_2$ as
\begin{equation}
W = P_1 (\phi_1^2 + M_1^2) + P_2 (\phi_2^2 + M_2^2) \;,
\end{equation}
where
\begin{equation}
M_1 =  \frac{h_2 M_A^2 - g_2 M_B^2}{g_1 h_2 - g_2 h_1} \quad \text{and}\quad M_2 =  \frac{h_1 M_A^2 - g_1 M_B^2}{g_2 h_1 - g_1 h_2} \;.
\end{equation}
Taking $P_1$ and $P_2$ as the new basis their $F$-term conditions fix the phases of the flavon vevs as discussed before.

In general the situation is even a bit more complicated. For example, for the $\phi_{123}$ flavon field at least two operators couple to the driving field $P_{123}$. If we have more operators than driving fields we cannot diagonalise the coupling matrix anymore. But we can redefine the driving fields appropriately and bring the coupling matrix to a triangular form. In this case we can apply an iterative procedure:
\begin{itemize}
\item We can start with the alignment of $\phi_{123}$, where we define a driving field $P_{123}$ coupling only to a combination of the operators as given in Eq.\ \eqref{eq:A4Phi123AlignmentA} or \eqref{eq:A4Phi123AlignmentB}, what we can always do as long as the coupling matrix is non-singular. After evaluating the $F$-term conditions of $P_{123}$ and choosing a vacuum, the value (including phases) of $\langle \phi_{123} \rangle$ (and $\langle \xi \rangle$) is fixed.
\item In the next iteration we redefine the driving fields in such a way that one driving field couples only to $\phi_{123}$ (and $\xi$) and another flavon, for example, $\phi_1$. Since the vev of $\phi_{123}$ (and $\xi$) is already fixed the new $F$-term condition again allows us to choose the value of the vev of $\phi_1$. The additional terms involving $\phi_{123}$ (and $\xi$) give only corrections to the mass parameter determining the mass scale of $\langle \phi_1 \rangle$.
\end{itemize}
This procedure can be iterated until all phases are fixed.

\section{$\boldsymbol{A_4}$ Messenger sector} \label{App:A4Messenger}

In this section we give the messenger sector for the $A_4$ model. The superpotential including messengers schematically looks like
\begin{equation}
\begin{split}
W &= M_{\Xi_i} \Xi_i \bar{\Xi}_i + M_{\Xi'_i} \Xi'_i \bar{\Xi}'_i  + M_{\Upsilon_{i;j}} \Upsilon_{i;j} \bar{\Upsilon}_{i;j}  + M_{\Theta_{i;j}} \Theta_{i;j} \bar{\Theta}_{i;j} \\
  & + M_{\Upsilon'} \Upsilon'_{123;123} \bar{\Upsilon}''_{123;123} + M_{\Upsilon''} \Upsilon''_{123;123} \bar{\Upsilon}'_{123;123}\\
  &+ \sum_{i,j,k} T_i \Xi_j \bar{H}_k + \sum_{i,j} N_i \Xi_j H + F \phi_i
\bar{\Xi}'_i  + H_{24} \bar{\Xi}_i \Xi'_i + H_{24} {\Xi}_i \bar\Xi'_i 
+ \sum_{i,j} N_i^2 \bar{\Upsilon}_j \\
  &+ \sum_{i,j,k,l} H \Theta_{i;j} \bar{\Upsilon}_{k,l} + \sum_{i,j} T_i T_j \bar{\Theta}_{i;j} + \sum_{i,j} \phi_i \phi_j \Upsilon_{i;j} + P_i \bar{\Upsilon}_{i;i}^2 + \Upsilon_{123} \xi^2 \;,
\end{split}
\end{equation}

\begin{table}
\centering
\begin{tabular}{cccccccccc} \toprule 
& $SU(5)$ & $A_4$   & $\mathbb{Z}^{(1)}_4$ & $\mathbb{Z}^{(2)}_4$ & $\mathbb{Z}^{(3)}_4$ & $\mathbb{Z}^{(4)}_4$ & $\mathbb{Z}^{(1)}_2$ & $\mathbb{Z}_2^{(2)}$ & $U(1)_R$ \\ \midrule
$\Xi_2$, $\bar{\Xi}_2$  & $\mathbf{5}$, $\mathbf{\overline{5}}$ & $\mathbf{1}$, $\mathbf{1}$ & 3, 1 & 3, 1 & 0, 0 & 0, 0 & 0, 0 & 0, 0 & 1, 1 \\
$\Xi_3$, $\bar{\Xi}_3$  & $\mathbf{5}$, $\mathbf{\overline{5}}$ & $\mathbf{1}$, $\mathbf{1}$ & 0, 0 & 0, 0 & 0, 0 & 0, 0 & 1, 1 & 0, 0 & 1, 1 \\
$\Xi_{123}$, $\bar{\Xi}_{123}$  & $\mathbf{5}$, $\mathbf{\overline{5}}$ & $\mathbf{1}$, $\mathbf{1}$ & 0, 0 & 0, 0 & 3, 1 & 0, 0 & 0, 0 & 0, 0 & 1, 1 \\
$\Xi_{23}$, $\bar{\Xi}_{23}$  & $\mathbf{5}$, $\mathbf{\overline{5}}$ & $\mathbf{1}$, $\mathbf{1}$ & 0, 0 & 0, 0 & 3, 1 & 3, 1 & 0, 0 & 0, 0 & 1, 1 \\ \midrule

$\Xi'_2$, $\bar{\Xi}'_2$  & $\mathbf{5}$, $\mathbf{\overline{5}}$ & $\mathbf{1}$, $\mathbf{1}$ & 3, 1 & 3, 1 & 0, 0 & 0, 0 & 0, 0 & 1, 1 & 1, 1 \\
$\Xi'_3$, $\bar{\Xi}'_3$  & $\mathbf{5}$, $\mathbf{\overline{5}}$ & $\mathbf{1}$, $\mathbf{1}$ & 0, 0 & 0, 0 & 0, 0 & 0, 0 & 1, 1 & 1, 1 & 1, 1 \\
$\Xi'_{123}$, $\bar{\Xi}'_{123}$  & $\mathbf{5}$, $\mathbf{\overline{5}}$ & $\mathbf{1}$, $\mathbf{1}$ & 0, 0 & 0, 0 & 3, 1 & 0, 0 & 0, 0 & 1, 1 & 1, 1 \\
$\Xi'_{23}$, $\bar{\Xi}'_{23}$  & $\mathbf{5}$, $\mathbf{\overline{5}}$ & $\mathbf{1}$, $\mathbf{1}$ & 0, 0 & 0, 0 & 3, 1 & 3, 1 & 0, 0 & 1, 1 & 1, 1 \\ \midrule

$\Upsilon_{1;1}$, $\bar{\Upsilon}_{1;1}$ & $\mathbf{1}$, $\mathbf{1}$ & $\mathbf{1}$, $\mathbf{1}$ & 2, 2 & 0, 0 & 0, 0 & 0, 0 & 0, 0 & 0, 0 & 2, 0 \\
$\Upsilon_{2;2}$, $\bar{\Upsilon}_{2;2}$ & $\mathbf{1}$, $\mathbf{1}$ & $\mathbf{1}$, $\mathbf{1}$ & 2, 2 & 2, 2 & 0, 0 & 0, 0 & 0, 0 & 0, 0 & 2, 0 \\
$\Upsilon_{23;23}$, $\bar{\Upsilon}_{23;23}$  & $\mathbf{1}$, $\mathbf{1}$ & $\mathbf{1}$, $\mathbf{1}$ & 0, 0 & 0, 0 & 2, 2 & 2, 2 & 0, 0 & 0, 0 & 2, 0 \\
$\Upsilon_{123;123}$, $\bar{\Upsilon}_{123;123}$  & $\mathbf{1}$, $\mathbf{1}$ & $\mathbf{1}$, $\mathbf{1}$ & 0, 0 & 0, 0 & 2, 2 & 0, 0 & 0, 0 & 0, 0 & 2, 0 \\
$\Upsilon'_{123;123}$, $\bar{\Upsilon}'_{123;123}$  & $\mathbf{1}$, $\mathbf{1}$ & $\mathbf{1}'$, $\mathbf{1}'$ & 0, 0 & 0, 0 & 2, 2 & 0, 0 & 0, 0 & 0, 0 & 2, 0 \\
$\Upsilon''_{123;123}$, $\bar{\Upsilon}''_{123;123}$  & $\mathbf{1}$, $\mathbf{1}$ & $\mathbf{1}''$, $\mathbf{1}''$ & 0, 0 & 0, 0 & 2, 2 & 0, 0 & 0, 0 & 0, 0 & 2, 0 \\ 
$\Upsilon_{1;3}$, $\bar{\Upsilon}_{1;3}$ & $\mathbf{1}$, $\mathbf{1}$ & $\mathbf{1}$, $\mathbf{1}$ & 1, 3 & 0, 0 & 0, 0 & 0, 0 & 1, 1 & 0, 0 & 2, 0 \\
$\Upsilon_{1;123}$, $\bar{\Upsilon}_{1;123}$ & $\mathbf{1}$, $\mathbf{1}$ & $\mathbf{1}$, $\mathbf{1}$ & 1, 3 & 0, 0 & 1, 3 & 0, 0 & 0, 0 & 0, 0 & 2, 0 \\
$\Upsilon_{3;123}$, $\bar{\Upsilon}_{3;123}$ & $\mathbf{1}$, $\mathbf{1}$ & $\mathbf{1}$, $\mathbf{1}$ & 0, 0 & 0, 0 & 1, 3 & 0, 0 & 1, 1 & 0, 0 & 2, 0 \\ \midrule

$\Theta_{1;1}$, $\bar{\Theta}_{1;1}$  & $\mathbf{5}$, $\mathbf{\overline{5}}$ & $\mathbf{1}$, $\mathbf{1}$ & 2, 2 & 0, 0 & 0, 0 & 0, 0 & 0, 0 & 0, 0 & 2, 0 \\
$\Theta_{2;2}$, $\bar{\Theta}_{2;2}$ & $\mathbf{5}$, $\mathbf{\overline{5}}$ & $\mathbf{1}$, $\mathbf{1}$ & 0, 0 & 0, 0 & 2, 2 & 0, 0 & 0, 0 & 0, 0 & 2, 0 \\
$\Theta_{1;2}$, $\bar{\Theta}_{1;2}$  & $\mathbf{5}$, $\mathbf{\overline{5}}$ & $\mathbf{1}$, $\mathbf{1}$ & 3, 1 & 0, 0 & 3, 1 & 0, 0 & 0, 0 & 0, 0 & 2, 0 \\
$\Theta_{1;3}$, $\bar{\Theta}_{1;3}$  & $\mathbf{5}$, $\mathbf{\overline{5}}$ & $\mathbf{1}$, $\mathbf{1}$ & 3, 1 & 0, 0 & 0, 0 & 0, 0 & 1, 1 & 0, 0 & 2, 0 \\
$\Theta_{2;3}$, $\bar{\Theta}_{2;3}$  & $\mathbf{5}$, $\mathbf{\overline{5}}$ & $\mathbf{1}$, $\mathbf{1}$ & 0, 0 & 0, 0 & 3, 1 & 0, 0 & 1, 1 & 0, 0 & 2, 0 \\  \bottomrule
\end{tabular}
\caption{List of messenger fields for the $A_4$ model, which give the desired terms in the superpotential after integrating them out. \label{tab:A4MessengerSector}} 
\end{table}

\noindent where the sum over the indices is taken over the fields listed in Tabs.\ \ref{tab:A4FlavonSector}, \ref{tab:A4MatterSector} and \ref{tab:A4MessengerSector} and the coefficients of the operators are dropped for the sake of simplicity. Note the mass terms for the primed $\Upsilon$ messengers. Due to our notation where the number of primes is the same as in the $A_4$ representation the mass terms are crossed. Keep here as well in mind that depending on the chosen option for the alignment of $\phi_{123}$ different messengers are present or not. To be more concrete the primed $\Upsilon$ messengers are present in option A and not present in option B. Although at the effective level option A seems to have much less fields than option B this is partially compensated at the messenger level.

\begin{figure}
\centering
\includegraphics[scale=0.68]{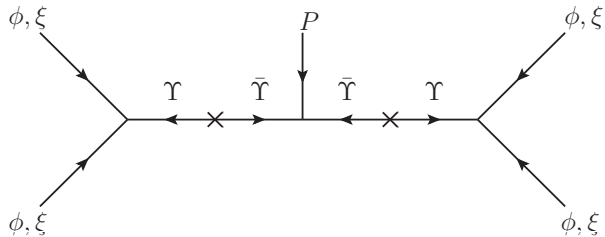}
\caption{Diagrams giving the non-renormalisable operators in the superpotential for the alignment of the flavon vevs. \label{fig:A4Alignment}}
\end{figure}

In Fig.\ \ref{fig:A4Alignment} we show the diagrams which give the non-renormalisable terms in the superpotential in Eqs.\ \eqref{eq:A4Phi123AlignmentA}-\eqref{eq:A4Phi23Alignment}. There is only one class of diagrams and one class of messengers involved. Due to the fact that every flavon $\phi$ has its own symmetries it is quite suggestive that these diagrams plus the renormalisable ones give the leading order operators.

\begin{figure}
\centering
\includegraphics[scale=0.68]{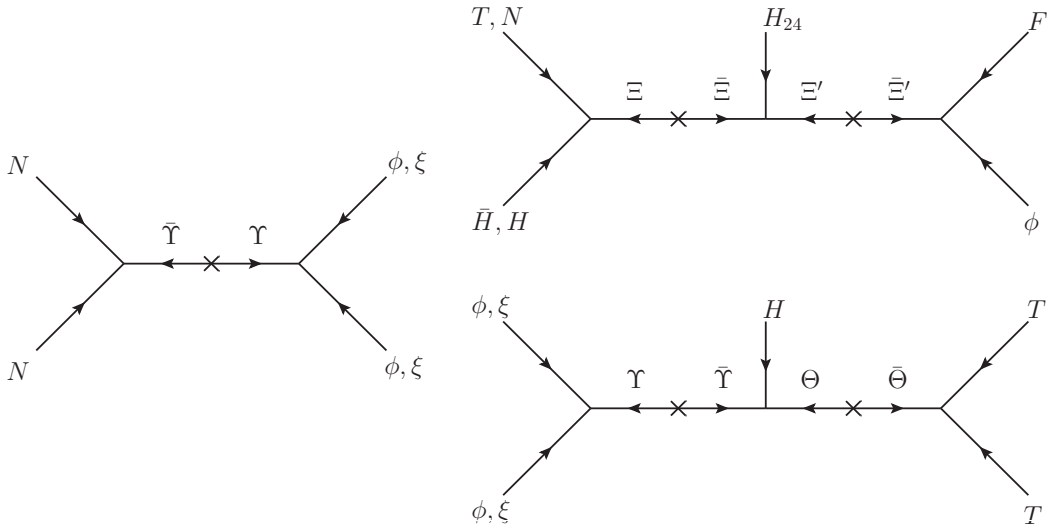}
\caption{Diagrams giving the non-renormalisable operators for the Yukawa couplings and right-handed neutrino masses. \label{fig:A4Yukawas}}
\end{figure}

In Fig.\ \ref{fig:A4Yukawas} we give the diagrams generating the Yukawa couplings and right-handed neutrino masses. The $\Upsilon$ messengers which already appeared in the diagrams for the flavon potential reappear here in the diagrams giving the right-handed neutrino masses and up-type quark Yukawa couplings.

\section{\label{app-messenger}A high energy completion of the $\boldsymbol{S_4}$ model}

This appendix presents the details of a possible high energy completion of our
$S_4$ model. The messenger sector is given in Tab.~\ref{tab:s4mess}. 
\begin{table}[p]
\centering
$$
\hspace{-0.4cm}\begin{array}{ccccccccccccc}\toprule
\text{Messengers} & SU(5) & S_4 & U(1)_R & {\mathbb{Z}}_4^{(1)} & {\mathbb{Z}}_4^{(2)} & {\mathbb{Z}}_2^{(3)} &
{\mathbb{Z}}_2^{(4)} & {\mathbb{Z}}_2^{(5)} & {\mathbb{Z}}_2^{(6)} & {\mathbb{Z}}_2^{(7)}  
& \wt {\mathbb{Z}}_2^{(k)} & \wt {\mathbb{Z}}_4^{(17)} \\\midrule
\Omega_1,\ol\Omega_1 & {\bf 5},{\bf \ol 5} & {\bf 2},{\bf 2} & 0,2 & 2,2 & 0,0 & 0,0 & 0,0 & 0,0 & 0,0
& 0,0 & 0,0 & 0,0 \\
\Omega_2,\ol\Omega_2 & {\bf{\ol{10}}},{\bf{10}} & {\bf 2},{\bf 2} & 1,1 & 0,0 & 0,0 & 1,1 & 1,1 & 0,0 & 0,0
& 0,0 & 0,0 & 0,0 \\
\Omega_3,\ol\Omega_3 & {\bf{\ol{10}}},{\bf{10}} & {\bf 3},{\bf 3} & 1,1 & 0,0 & 1,3 & 0,0 & 1,1 & 0,0 & 0,0
& 1,1 & 0,0 & 0,0 \\
\Omega_4,\ol\Omega_4 & {\bf 5},{\bf \ol 5} & {\bf 1},{\bf 1} & 0,2 & 0,0 & 2,2 & 0,0 & 0,0 & 0,0 & 0,0
& 0,0 & 0,0 & 0,0 \\
\Omega_5,\ol\Omega_5 & {\bf \ol 5},{\bf 5} & {\bf 3},{\bf 3} & 0,2 & 0,0 & 3,1 & 0,0 & 0,0 & 0,0 & 0,0
& 0,0 & 0,0 & 0,0  \\
\Omega_6,\ol\Omega_6 & {\bf 5},{\bf \ol 5} & {\bf 1},{\bf 1} & 1,1 & 3,1 & 3,1 & 0,0 & 0,0 & 1,1 & 0,0
& 0,0 & 0,0 & 0,0  \\
\Omega_7,\ol\Omega_7 & {\bf{\ol{45}}},{\bf{45}} & {\bf 2},{\bf 2} & 0,2 & 0,0 & 3,1 & 0,0 & 1,1 & 1,1
& 0,0 & 0,0 & 0,0 & 0,0 \\
\Omega_8,\ol\Omega_8 & {\bf 5},{\bf \ol 5} & {\bf 2},{\bf 2} & 1,1 & 3,1 & 0,0 & 0,0 & 0,0 & 0,0 & 0,0
& 1,1 & 0,0 & 0,0  \\
\Omega_9,\ol\Omega_9 & {\bf 5},{\bf \ol 5} & {\bf 2},{\bf 2} & 1,1 & 2,2 & 0,0 & 0,0 & 1,1 & 0,0 & 0,0
& 1,1 & 0,0 & 0,0  \\\midrule
\Sigma_1,\ol\Sigma_1 & {\bf 1},{\bf 1} & {\bf 3'},{\bf 3'} & 0,2 & 0,0 & 2,2 & 0,0 & 0,0 &
0,0 & 0,0 & 0,0 & \delta_{7k},\delta_{7k} & 0,0 \\
 \Sigma_2,\ol\Sigma_2 & {\bf 1},{\bf 1} & {\bf 2},{\bf 2} & 0,2 & 0,0 & 2,2 & 0,0 & 0,0 & 0,0
& 0,0 & 0,0 & \delta_{7k},\delta_{7k} & 0,0 \\
 \Sigma_3,\ol\Sigma_3 & {\bf 1},{\bf 1} & {\bf 3'},{\bf 3'} & 0,2 & 0,0 & 2,2 & 0,0 & 0,0 &
0,0 & 0,0 & 0,0 & \delta_{8k},\delta_{8k} & 0,0 \\
 \Sigma_4,\ol\Sigma_4 & {\bf 1},{\bf 1} & {\bf 2},{\bf 2} & 0,2 & 1,3 & 3,1 & 0,0 & 0,0 & 0,0
& 1,1 & 0,0 & \delta_{1k},\delta_{1k} & 0,0 \\
 \Sigma_5,\ol\Sigma_5 & {\bf 1},{\bf 1} & {\bf 3},{\bf 3} & 0,2 & 3,1 & 1,3 & 0,0 & 0,0 & 0,0
& 0,0 & 1,1 & \delta_{2k},\delta_{2k} & 0,0 \\
 \Sigma_6,\ol\Sigma_6 & {\bf 1},{\bf 1} & {\bf 3'},{\bf 3'} & 0,2 & 3,1 & 1,3 & 0,0 & 0,0 &
0,0 & 0,0 & 1,1 & \delta_{2k},\delta_{2k} & 0,0 \\
 \Sigma_7,\ol\Sigma_7 & {\bf 1},{\bf 1} & {\bf 2},{\bf 2} & 0,2 & 2,2 & 2,2 & 0,0 & 0,0 & 0,0
& 0,0 & 0,0 & \delta_{2k},\delta_{2k} & 0,0 \\
 \Sigma_8,\ol\Sigma_8 & {\bf 1},{\bf 1} & {\bf 2},{\bf 2} & 0,2 & 1,3 & 3,1 & 0,0 & 0,0 & 0,0
& 1,1 & 0,0 & \delta_{2k},\delta_{2k} & 0,0 \\
 \Sigma_9,\ol\Sigma_9 & {\bf 1},{\bf 1} & {\bf 2},{\bf 2} & 0,2 & 2,2 & 3,1 & 0,0 & 0,0 & 1,1
& 0,0 & 1,1 & \delta_{3k},\delta_{3k} & 0,0 \\
 \Sigma_{10},\ol\Sigma_{10} & {\bf 1},{\bf 1} & {\bf 3},{\bf 3} & 0,2 & 1,3 & 0,0 & 0,0 & 0,0 &
1,1 & 0,0 & 0,0 & \delta_{3k},\delta_{3k} & 0,0 \\
\Sigma_{11},\ol\Sigma_{11} & {\bf 1},{\bf 1} & {\bf 3},{\bf 3} & 0,2 & 3,1 & 2,2 & 0,0 & 0,0 &
1,1 & 0,0 & 0,0 & \delta_{4k},\delta_{4k} & 0,0 \\
\Sigma_{12},\ol\Sigma_{12} & {\bf 1},{\bf 1} & {\bf 3'},{\bf 3'} & 0,2 & 2,2 & 2,2 & 0,0 & 0,0
& 0,0 & 0,0 & 0,0 & \delta_{4k},\delta_{4k} & 0,0 \\
\Sigma_{13},\ol\Sigma_{13} & {\bf 1},{\bf 1} & {\bf 3},{\bf 3} & 0,2 & 1,3 & 3,1 & 0,0 & 0,0 &
0,0 & 0,0 & 1,1 & \delta_{4k},\delta_{4k} & 0,0 \\
\Sigma_{14},\ol\Sigma_{14} & {\bf 1},{\bf 1} & {\bf 2},{\bf 2} & 0,2 & 2,2 & 0,0 & 0,0 & 0,0 &
0,0 & 0,0 & 0,0 & \delta_{5k},\delta_{5k} & 0,0 \\
\Sigma_{15},\ol\Sigma_{15} & {\bf 1},{\bf 1} & {\bf 2},{\bf 2} & 0,2 & 1,3 & 0,0 & 0,0 & 1,1 &
0,0 & 0,0 & 0,0 & \delta_{6k},\delta_{6k} & 0,0 \\
\Sigma_{16},\ol\Sigma_{16} & {\bf 1},{\bf 1} & {\bf 1},{\bf 1} & 0,2 & 2,2 & 0,0 & 0,0 & 0,0 &
0,0 & 0,0 & 0,0 & \delta_{9k},\delta_{9k} & 0,0 \\
\Sigma_{17},\ol\Sigma_{17} & {\bf 1},{\bf 1} & {\bf 1'},{\bf 1'} & 0,2 & 2,2 & 0,0 & 0,0 & 0,0
& 0,0 & 0,0 & 0,0 & \delta_{10\,k},\delta_{10\,k} & 0,0 \\
\Sigma_{18},\ol\Sigma_{18} & {\bf 1},{\bf 1} & {\bf 1},{\bf 1} & 0,2 & 0,0 & 2,2 & 0,0 & 0,0 &
0,0 & 0,0 & 0,0 & \delta_{11\,k},\delta_{11\,k} & 0,0 \\
\Sigma_{19},\ol\Sigma_{19} & {\bf 1},{\bf 1} & {\bf 1'},{\bf 1'} & 0,2 & 1,3 & 0,0 & 1,1 & 0,0
& 0,0 & 0,0 & 0,0 & \delta_{12\,k},\delta_{12\,k} & 0,0 \\
\Sigma_{20},\ol\Sigma_{20} & {\bf 1},{\bf 1} & {\bf 2},{\bf 2} & 0,2 & 1,3 & 0,0 & 0,0 & 1,1 &
0,0 & 0,0 & 0,0 & \delta_{13\,k},\delta_{13\,k} & 0,0 \\
\Sigma_{21},\ol\Sigma_{21} & {\bf 1},{\bf 1} & {\bf 3},{\bf 3} & 0,2 & 1,3 & 0,0 & 0,0 & 0,0 &
1,1 & 0,0 & 0,0 & \delta_{14\,k},\delta_{14\,k} & 0,0 \\
\Sigma_{22},\ol\Sigma_{22} & {\bf 1},{\bf 1} & {\bf 2},{\bf 2} & 0,2 & 1,3 & 1,3 & 0,0 & 0,0 &
0,0 & 1,1 & 0,0 & \delta_{15\,k},\delta_{15\,k} & 0,0 \\
\Sigma_{23},\ol\Sigma_{23} & {\bf 1},{\bf 1} & {\bf 2},{\bf 2} & 0,2 & 0,0 & 2,2 & 0,0 & 0,0 &
0,0 & 0,0 & 0,0 & \delta_{15\,k},\delta_{15\,k} & 0,0 \\
\Sigma_{24},\ol\Sigma_{24} & {\bf 1},{\bf 1} & {\bf 3'},{\bf 3'} & 0,2 & 1,3 & 1,3 & 0,0 & 0,0
& 0,0 & 0,0 & 1,1 & \delta_{16\,k},\delta_{16\,k} & 0,0 \\
\Sigma_{25},\ol\Sigma_{25} & {\bf 1},{\bf 1} & {\bf 2},{\bf 2} & 0,2 & 0,0 & 2,2 & 0,0 & 0,0 &
0,0 & 0,0 & 0,0 & \delta_{16\,k},\delta_{16\,k} & 0,0 \\
\Sigma_{26},\ol\Sigma_{26} & {\bf 1},{\bf 1} & {\bf 2},{\bf 2} & 0,2 & 0,0 & 0,0 & 0,0 & 0,0 &
0,0 & 0,0 & 0,0 & 0,0 & 3,1  \\
\Sigma_{27},\ol\Sigma_{27} & {\bf 1},{\bf 1} & {\bf 1},{\bf 1} & 0,2 & 3,1 & 1,3 & 0,0 & 0,0 &
0,0 & 1,1 & 0,0 &  0,0 & 3,1 \\
\Sigma_{28},\ol\Sigma_{28} & {\bf 1},{\bf 1} & {\bf 1},{\bf 1} & 0,2 & 3,1 & 1,3 & 0,0 & 0,0 &
0,0 & 1,1 & 0,0 &  0,0 & 2,2  \\
\Sigma_{29},\ol\Sigma_{29} & {\bf 1},{\bf 1} & {\bf 2},{\bf 2} & 0,2 & 2,2 & 2,2 & 0,0 & 0,0 &
0,0 & 0,0 & 0,0 &  0,0 & 2,2  \\
\Sigma_{30},\ol\Sigma_{30} & {\bf 1},{\bf 1} & {\bf 2},{\bf 2} & 0,2 & 1,3 & 3,1 & 0,0 & 0,0 &
0,0 & 1,1 & 0,0 &  0,0 & 2,2  \\
\Sigma_{31},\ol\Sigma_{31} & {\bf 1},{\bf 1} & {\bf 1'},{\bf 1'} & 0,2 & 2,2 & 0,0 & 0,0 & 0,0
& 0,0 & 0,0 & 0,0 &  0,0 & 1,3 \\\bottomrule
\end{array}
$$
\caption{\label{tab:s4mess}The list of messengers in the $S_4$ model.}
\end{table}
With this set of messengers, the renormalisable superpotential including matter,
Higgs, flavon, driving and messenger fields can be worked out
straightforwardly and takes the following form
\be
W^{\mathrm{ren.}} ~=~ W^{\mathrm{ren.}}_{\mathrm{Yuk}} ~+~
W^{\mathrm{ren.}}_{\mathrm{flavon}}\ ,
\ee
with
\bea
W^{\mathrm{ren.}}_{\mathrm{Yuk}} &=&
FNH_5 + N (\phi^\nu_{3'}+\phi^\nu_2+\phi^\nu_{1}) N + T_3T_3H_5 \notag\\
&&+~ TT\Omega_1  + H_5 \phi^u_2 \ol \Omega_1 \notag\\
&&+~ T\phi^u_{1'} \Omega_2 + H_5 \ol \Omega_2 \ol\Omega_2\notag\\
&&+~ T\phi^d_3 \Omega_3 + H_5 \phi^\nu_1 \ol\Omega_4 + \ol\Omega_3\ol\Omega_3
\Omega_4 \notag\\
&&+~ T_3 F \Omega_5 + H_{\ol 5} \phi^d_3 \ol\Omega_5 \notag\\
&&+~ T \ol \Omega_6 \Omega_7 + F\wt\phi^d_3 \Omega_6 + H_{\ol{45}} \phi^d_2
\ol\Omega_7 \notag\\
&&+~ TH_{\ol 5} \ol\Omega_9 + F\phi^d_3 \Omega_8 + \wt\phi^d_2
\ol\Omega_8\Omega_9 \notag\\
&&+~M \sum_{i=1}^9 \Omega_i \ol \Omega_i \ ,
\eea
\bea
W^{\mathrm{ren.}}_{\mathrm{flavon}} &=&
Y^\nu_2 \phi^\nu_{3'} \Sigma_1 + 
\zeta^{Y^\nu_2}_1 \phi^\nu_{3'}\ol\Sigma_1 \notag\\
&&+~ Y^\nu_2 (\phi^\nu_{1}+\phi^\nu_{2}) \Sigma_2 + 
\zeta^{Y^\nu_2}_1 \phi^\nu_{2}\ol\Sigma_2 \notag\\
&&+~ Z^\nu_{3'} (\phi^\nu_{1}+\phi^\nu_{2}+\phi^\nu_{3'}) \Sigma_3 +
\zeta^{Z^\nu_{3'}}_1 \phi^\nu_{3'} \ol\Sigma_3\notag\\
&&+~ X^d_1\phi^d_2  \Sigma_4+
\zeta^{X^d_1}_1\phi^d_2  \ol\Sigma_4\notag\\
&&+~ Y^d_2 \phi^d_3(\Sigma_5+\Sigma_6)+
\phi^d_3 (\ol\Sigma_5+\ol\Sigma_6) \Sigma_7+
\phi^d_2 \ol\Sigma_7 \Sigma_8+
\zeta^{Y^d_2}_1\phi^d_2\ol\Sigma_8\notag\\
&&+~\wt X^d_1\phi^d_2\Sigma_9 +
\phi^d_3\ol\Sigma_9 \Sigma_{10}+
\zeta^{\wt X^d_1}_1\wt\phi^d_3\ol\Sigma_{10}\notag\\
&&+~\wt X^{\nu d}_{1'}\phi^\nu_{3'}\Sigma_{11}+
\wt\phi^d_3 \ol\Sigma_{11}\Sigma_{12}+
\phi^d_3 \ol\Sigma_{12}\Sigma_{13}+
\zeta^{\wt X^{\nu d}_{1'}}_1\phi^d_3 \ol\Sigma_{13}\notag\\
&&+~Y^{du}_2\phi^d_2\Sigma_{14}+
\zeta^{Y^{du}_2}_1\phi^u_2 \ol\Sigma_{14}\notag\\
&&+~X^{\nu d}_{1'}\phi^\nu_{2}\Sigma_{15}+
\zeta^{X^{\nu d}_{1'}}_1\wt\phi^d_2\ol\Sigma_{15}\notag\\
&&+~P_0^{(1)}\xi_1\Sigma_{16}+
\zeta^{P_0^{(1)}}_1\xi_1\ol\Sigma_{16}\notag\\
&&+~P_0^{(2)}\wt\xi_{1'}\Sigma_{17}+
\zeta^{P_0^{(2)}}_1\wt\xi_{1'}\ol\Sigma_{17}\notag\\
&&+~P_0^{(3)}\phi^\nu_{1}\Sigma_{18}+
\zeta^{P_0^{(3)}}_1\phi^\nu_{1}\ol\Sigma_{18}\notag\\
&&+~P_1^{(1)}\phi^u_{1'}\Sigma_{19}+
\zeta^{P_1^{(1)}}_1\phi^u_{1'}\ol\Sigma_{19}\notag\\
&&+~P_1^{(2)}\wt\phi^d_2\Sigma_{20}+
\zeta^{P_1^{(2)}}_1\wt\phi^d_2\ol\Sigma_{20}\notag\\
&&+~P_1^{(3)}\wt\phi^d_3  \Sigma_{21}+
\zeta^{P_1^{(3)}}_1\wt\phi^d_3 \ol \Sigma_{21}\notag\\
&&+~P_1^{(4)} \phi^d_2  \Sigma_{22}+
\phi^d_2  \ol\Sigma_{22}\Sigma_{23}+
\zeta^{P_1^{(4)}}_1\phi^\nu_{2}\ol\Sigma_{23}\notag\\
&&+~\wt P_{1'}^{(1)}\phi^d_3\Sigma_{24}+
\phi^d_3\ol\Sigma_{24}\Sigma_{25}+
\zeta^{\wt P_{1'}^{(1)}}_1\phi^\nu_{2}\ol\Sigma_{25}\notag\\
&&+~\wt P_{1'}^{(2)}\phi^u_2\Sigma_{26}+
\phi^d_2\ol\Sigma_{26}\Sigma_{27}+
\wt\zeta^{\wt P_{1'}^{(2)}}_1\ol\Sigma_{27}\Sigma_{28}+
\phi^d_2\ol\Sigma_{28}\Sigma_{29}+
\phi^d_2\ol\Sigma_{29}\Sigma_{30}+
\zeta^{\wt P_{1'}^{(2)}}_1\phi^d_2\ol\Sigma_{30}~~~~\notag\\
&&+~\wt P_{1'}^{(2)}\zeta^{\wt P_{1'}^{(2)}}_1\Sigma_{31}+
\wt\zeta^{\wt P_{1'}^{(2)}}_1\wt\xi_{1'}\ol\Sigma_{31}\notag\\
&&+~ \sum_{i=1}^3 m^{(i)} P_0^{(i)} \zeta^{P_0^{(i)}}_1  + 
\sum_{i=1}^4 P_1^{(i)} \zeta^{P_1^{(i)}}_1\xi_1+ 
\wt P_{1'}^{(1)} \zeta_1^{\wt P_{1'}^{(1)}} \wt \xi_{1'}+
M \sum_{i=1}^{31} \Sigma_i \ol\Sigma_i \ .
\eea
These operators are grouped such that, after integrating out the messenger
fields, each line gives rise to one particular non-renormalisable term in the
effective superpotential, i.e. Eqs.~(\ref{up1}-\ref{nu2}) and the terms labelled~(\ref{fla8}-\ref{fla7},\ref{real1new}-\ref{real3new}).

\end{appendix}

\end{document}